\documentclass{aa}  

\usepackage{graphicx}
\usepackage{txfonts}
\usepackage{natbib}
\usepackage{aalongtable}

\begin{document} 

\title{Supersaturation and activity-rotation relation in PMS stars: \\ the young cluster h~Per\thanks{Tables 1 and 2 Tables are available in electronic form at the CDS via anonymous ftp to cdsarc.u-strasbg.fr (130.79.128.5) or via http://cdsweb.u-strasbg.fr/cgi-bin/qcat?J/A+A/...}}

\titlerunning{Supersaturation, activity and rotation: the young cluster h~Per}

\author{C. Argiroffi\inst{1,2}, M. Caramazza\inst{2}, G. Micela\inst{2}, S. Sciortino\inst{2}, E. Moraux\inst{3}, J. Bouvier\inst{3}, \and E. Flaccomio\inst{2}}
\authorrunning{C. Argiroffi et al.}

\institute{
Dip. di Fisica e Chimica, Universit\`a di Palermo, Piazza del Parlamento 1, 90134, Palermo, Italy, \email{argi@astropa.unipa.it}
\and
INAF - Osservatorio Astronomico di Palermo, Piazza del Parlamento 1, 90134, Palermo, Italy,
\and
Univ. Grenoble Alpes, IPAG, F-38000 Grenoble, France \\
CNRS, IPAG, F-38000 Grenoble, France}
\date{Received May 18, 2015; accepted February 10, 2016}

\abstract
{Several studies showed that the magnetic activity of late-type main-sequence (MS) stars is characterized by different regimes and that their activity levels are well described by the Rossby number, $Ro$, defined as the ratio between the rotational period $P_{\rm rot}$ and the convective turnover time. Very young pre-main-sequence (PMS) stars show, similarly to MS stars, intense magnetic activity. However, they do not show clear activity-rotation trends, and it still debated which stellar parameters determine their magnetic activity levels.}
{To bridge the gap between MS and PMS stars, we studied the activity-rotation relation in the young cluster \object{h~Persei}, a $\sim13$\,Myr old cluster, that contains both fast and slow rotators. The cluster members have ended their accretion phase and have developed a radiative core. It therefore offers us the opportunity of studying the activity level of intermediate-age PMS stars with different rotational velocities, excluding any interactions with the circumstellar environment.}
{We constrained the magnetic activity levels of h~Per members by measuring their X-ray emission from a {\it Chandra} observation, while rotational periods were obtained previously in the framework of the MONITOR project. By cross-correlating these data, we collected a final catalog of 414 h~Per members with known rotational period, effective temperature, and mass. In 169 of these, X-ray emission has also been detected.}
{We found that h~Per members with $1.0\,M_{\odot}<M_{\star}<1.4\,M_{\odot}$ display different activity regimes: fast rotators clearly show supersaturation, while slower rotators have activity levels compatible to the non-saturated regime. At 13\,Myr, h~Per is therefore the youngest cluster showing activity-rotation regimes analogous to those of MS stars, indicating that at this age, magnetic field production is most likely regulated by the $\alpha\Omega$ type dynamo. Moreover, we observed that supersaturation is better described by $P_{\rm rot}$ than $Ro$, and that the observed patterns are compatible with the hypothesis of centrifugal stripping. In this scenario we inferred that coronae can produce structures as large as $\sim2\,R_{\star}$ above the stellar surface.}
{}

\maketitle

\section{Introduction}


Main-sequence (MS) late-type stars, including the Sun, produce intense magnetic fields, as evidenced by many observational features: photometric variability that is due to spots, enhanced chromospheric lines, frequent flaring activity, and intense X-ray emission \citep[e.g.,][]{Berdyugina2005,KovariOlah2014}. In these stars X-rays are emitted by the stellar coronae, the outer stellar atmosphere where hot plasma is confined and heated by the stellar magnetic field \citep[e.g.,][]{FavataMicela2003,Gudel2004}. This makes stellar X-ray emission one of the best probes of stellar magnetic activity.


Stellar magnetic fields are thought to be produced by dynamo processes that in turn are caused by plasma motions in the stellar interior. The role of stellar rotation and how stellar activity increases with stellar rotational velocity was initially described by \citet{Skumanich1972} and \citet{PallaviciniGolub1981}. Then \citet{NoyesHartmann1984} also included the role of convective motions and proved that stellar activity levels do not depend on rotation alone, but are indeed better described by the Rossby number, $Ro$, defined as the ratio between the rotational period $P_{\rm rot}$ and the convective turnover time $\tau$, which is the characteristic time taken by the plasma to cover a given distance in the convective envelope. Considering these results and what is known for the Sun, it is believed that all the stars with an inner radiative core and an outer convective envelope develop a $\alpha\Omega$ type dynamo, originating in a thin shell named tachocline at the interface between these two regions \citep{Parker1955,SpiegelWeiss1980}.


Several studies based on large samples of late-type MS stars showed that stellar activity is characterized by different regimes \citep[e.g.,][]{DobsonRadick1989,PizzolatoMaggio2003,WrightDrake2011}. In the non-saturated regime, that is, for $Ro>0.13$, the stellar X-ray luminosity anticorrelates with $Ro$, with the fractional X-ray emission $L_{\rm X}/L_{\rm bol}$ scaling as $Ro^{-2.7}$ \citep{WrightDrake2011}. This non-saturated regime, showing how activity levels and internal motions are linked, shows that stellar activity is indeed produced by a dynamo mechanism. For increasing rotational velocities, or more precisely, for $Ro<0.13$, the $L_{\rm X}/L_{\rm bol}$ of MS stars saturates to its maximum level, which is $L_{\rm X}/L_{\rm bol}\approx10^{-3}$, defining the so-called saturated regime. It is still debated whether this saturated level is due to an intrinsic saturation of the dynamo efficiency or to external constraints, like the full coverage of stellar surface with active regions. In addition to non-saturation and saturation, a few studies suggested that in very rapidly rotating MS stars, a third regime probably occurs. These very rapid rotators show $L_{\rm X}/L_{\rm bol}$ ratios lower than the saturated level \citep{RandichSchmitt1996,ProsserRandich1996,JamesJardine2000,JeffriesJackson2011}, which indicates the existence of a third regime, called supersaturation. This behavior was observed only for a very few stars belonging to young clusters ($\sim30-50$\, Myr), probably because of the decreasing stellar rotational
velocities in older clusters.


Similarly to what occurs in MS stars, late-type pre-main-sequence (PMS) stars are also magnetically active, producing strong magnetic fields and manifesting intense coronal emission. Several studies investigated the properties of coronal activity of PMS stars, focusing almost entirely on very young ($\sim1-3$\,Myr) clusters. These studies found that in PMS stars, differently than in MS stars,  $L_{\rm X}/L_{\rm bol}$ and $Ro$ are not correlated, with $L_{\rm X}/L_{\rm bol}$ showing a very large scatter \citep[e.g.,][]{StassunArdila2004,PreibischKim2005,BriggsGudel2007}. The activity level of PMS stars, always showing a huge scatter, correlates with the accretion status, with accreting stars displaying on average lower $L_{\rm X}/L_{\rm bol}$ than non-accreting stars \citep[e.g.,][]{FlaccomioMicela2003,PreibischKim2005}. It is still debated which mechanism causes this difference \citep[e.g.,][]{PreibischKim2005,TelleschiGudel2007,FlaccomioMicela2010,FlaccomioMicela2012}.
Some studies moreover also observed a positive correlation between $L_{\rm X}/L_{\rm bol}$ and the stellar rotational period \citep{FeigelsonGaffney2003,StassunArdila2004,PreibischKim2005,HendersonStassun2012}, a trend analogous the supersaturation phenomenon observed in young MS stars. However, this trend is not ubiquitous \citep{RebullStauffer2006,AlexanderPreibisch2012}. These results indicate that it is still unclear which stellar parameters determine the magnetic activity levels in PMS stars. Identifying this would help in constraining the physical mechanism causing or regulating their magnetic activity.


Diverse mechanisms could generate the different magnetic properties between MS and PMS stars. First, PMS and MS stars have different internal structures: PMS stars have deeper and more massive convective envelopes than MS stars, with very young stars (with mass of up to $\sim1.0\,M_{\odot}$) being even fully convective. Moreover, considering that the different layers of internal stellar structure may have different rotational velocity, and that these velocities evolve on different timescales \citep{GalletBouvier2013}, then the different magnetic properties of PMS and MS stars could also be related to a different internal distribution of rotational velocities. Finally it is worth noting that PMS stars, especially at very young ages, still accrete material from their circumstellar disks, experiencing exchange of mass, energy, and angular momentum. The accretion process, braking or affecting stellar rotation or interacting with the stellar magnetosphere, might affect the magnetic activity of accreting stars.


To bridge the gap between the well-constrained case of MS stars and the puzzling case of very young PMS stars, we studied the activity-rotation relation in the young cluster \object{h~Per}, which is a rich cluster, $\sim13$\,Myr old, located at 2290\,pc, and characterized by a moderate interstellar absorption ($E(B-V)\sim0.55$, which corresponds to $A_{\rm V}=1.7$). Because of its age, the h~Per cluster offers several advantages: {\it a)} it contains both fast and slow rotators, allowing us therefore to test the different regimes of stellar dynamo, and in particular to search for and investigate the supersaturation phenomenon; {\it b)} accretion processes are completed, allowing us therefore to test the stellar magnetic activity excluding any interactions with the circumstellar environment; {\it c)} all the stars with $0.5\,{\rm M_{\odot}}<M<1.5\,{\rm M_{\odot}}$ at the h~Per age have already developed a radiative core and still preserve a convective envelope, which means that they have an internal structure similar to that of slightly older MS stars, where the $\alpha\Omega$ dynamo is already at work. Investigating the case of a young cluster also allows us to constrain whether stellar activity depends on Rossby number alone, as for MS stars, or whether it behaves differently,
which would indicate that magnetic activity also depends on stellar evolutionary phase, hence on mass and age.

This paper is organized as follows: observation properties, data analysis, source identification, and parameter determination are reported in Sect.~\ref{data}; properties of the selected sample of h~Per members are presented in Sect.~\ref{sampleprop}; Sect.~\ref{activityanalysis} describes the search for activity-rotation relation, and the results are discussed in Sect.~\ref{disc}.

\begin{figure}
\centering
\includegraphics[width=8.5cm]{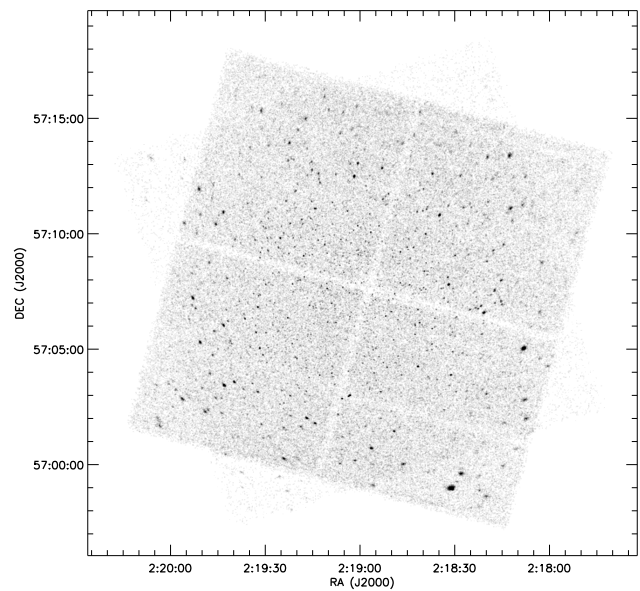}
\caption{ $17\arcmin\times17\arcmin$ field of view of the Chandra/ACIS-I observation of h~Per. In this image we detected 1002 X-ray sources.}
\label{plotCHANDRAfov}
\end{figure}

\section{Data analysis}
\label{data}

To investigate the activity-rotation relation in the h~Per cluster, we constrained the magnetic activity levels of h~Per members from a deep {\it Chandra} observation. Rotational periods were obtained by \citet{MorauxArtemenko2013} in the framework of the MONITOR project.

\subsection{X-ray source detection}

We obtained a deep {\it Chandra}/ACIS-I observation of the h~Per cluster in 2009. This observation has an exposure time of $189.8$\,ks, it is divided into three observing segments (Obs ID 09912, 09913, and 12021, PI G. Micela), and it is centered on $RA=02^{\rm h}\,19^{\rm m}\,02\fs20$ and $DEC=+57\degr07\arcmin12\farcs00$. Data reduction was performed in a standard way, using the CIAO~4.1 package and following the threads provided by the {\it Chandra} X-ray Center. X-ray data were finally filtered considering only events with energy ranging between 0.5 and 8.0\,keV.

We searched for X-ray sources using the {\tt PWDetect} code \citep{DamianiMaggio1997a,DamianiMaggio1997b}, a wavelet-based detection algorithm. We set the significance threshold to $4.7\sigma$, which, considering the background level of our observation, corresponds to an expected number of spurious detections of ten sources. For each detected source this code provides in addition to the source position and detection significance
 the background-subtracted count rate in the $0.5-8$\,keV band. We applied the {\tt PWDetect} code to the superposition of the three observing segments, shown in Fig.~\ref{plotCHANDRAfov}, collecting a list of 1010 X-ray sources. After a careful inspection, we rejected eight entries, corresponding to sources detected twice, obtaining therefore a final catalog of 1002 distinct X-ray sources, whose position and X-ray properties are reported in Table~\ref{tab:xsrc}. The sensitivity of our X-ray survey is not uniform over the {\it Chandra} field of view (fov): it is higher near the telescope axis, where the weakest X-ray sources detected have fluxes of $\sim10^{-7}\,{\rm ph\,s^{-1}\,cm^{-2}}$, while it diminishes in the outer regions (off axis $\sim10\arcmin$), where the limiting flux is $\sim5\times10^{-7}\,{\rm ph\,s^{-1}\,cm^{-2}}$.

\begin{figure}
\centering
\includegraphics[width=8.5cm]{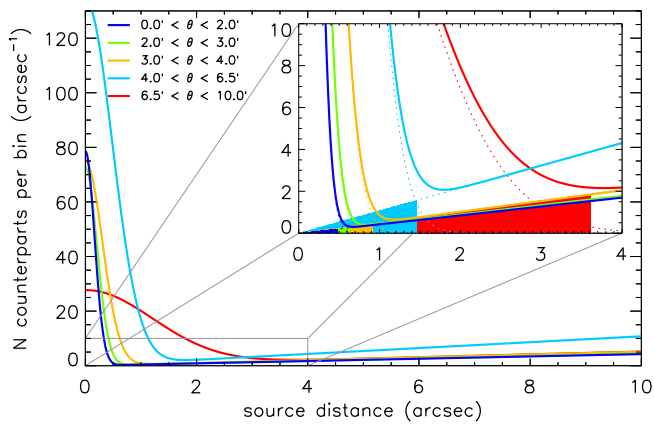}
\caption{Best-fit functions obtained from the distance distributions between X-ray sources and \citeauthor{MorauxArtemenko2013} catalog counterparts. We separately computed the distributions corresponding to X-ray sources located at different off-axis angle $\theta$ in the {\it Chandra} fov. The two components of each best-fit function, the Gaussian and the linear function, are indicated with dotted lines in the inset plot. The areas of the solid triangles indicate the expected number of spurious identification, assuming
a matching radius of $3\sigma$ for
each off-axis interval.}
\label{spurid}
\end{figure}

\subsection{X-ray source identification}

To identify the detected X-ray sources, we first cross-correlated them with the catalog presented by \citet{MorauxArtemenko2013} that contains 586 h~Per members with a measured rotational period, 541 of which fall within the {\it Chandra} fov. For a correct comparison of the two catalogs we corrected the X-ray positions for a systematic offset, with respect to the \citeauthor{MorauxArtemenko2013} catalog, of $\Delta\,RA = -0.28\arcsec$ and $\Delta\,DEC = 0.28\arcsec$. To infer the best-matching radius for the source identification, we constructed and inspected the distribution of distances of \citeauthor{MorauxArtemenko2013} sources with respect to X-ray sources. This distribution is well described by a Gaussian, centered on a distance of 0\arcsec, plus a linear function. These two components represent correlated and spurious identifications, respectively. This distribution is not the same over the {\it Chandra} fov, in particular, the Gaussian function becomes broader for increasing off-axis angle because of the decreasing accuracy of the X-ray position. We show in Fig.~\ref{spurid} the best-fit functions that describe these distributions, computed for different off-axis angle intervals. Considering these Gaussian functions, we adopted a matching radius of $3\sigma$, which corresponds to $\sim0.5\arcsec$ for sources near the telescope axis, and increases up to $\sim3.6\arcsec$ for large off-axis angle. From this comparison with the \citeauthor{MorauxArtemenko2013} catalog, we found h~Per member counterparts for 201 X-ray sources (with one X-ray source, X source ID = 187, being associated to two different counterparts). Considering the best-fit components describing the spurious associations (see Fig.~\ref{spurid}), and integrating them up to the assumed matching radii, we expect five spurious identifications. 384 h~Per members of the \citeauthor{MorauxArtemenko2013} catalog were left with no X-ray information. Since 339 of them are located within the {\it Chandra} fov, we estimated their upper limits in the X-ray band using the {\tt PWDetect} code.

To search for other h~Per members among the X-ray sources with no counterparts in the \citeauthor{MorauxArtemenko2013} catalog, we cross-correlated them with the h~Per member list presented by \citet[][Table~6]{CurrieHernandez2010}. Again we corrected the X-ray source positions for a systematic offset ($\Delta\,RA = -0.43\arcsec$ and $\Delta\,DEC = 0.07\arcsec$) emerged from a first comparison with the source position of \citet{CurrieHernandez2010}. We computed and investigated the distribution of distances between X-ray and optical sources, applying the same procedure as was
adopted for the counterparts search in the \citeauthor{MorauxArtemenko2013} catalog. In this case, the adopted matching radius ranges from $\sim0.6\arcsec$ to $\sim3.0\arcsec$ (corresponding to $3\sigma$  in this case as well) for increasing off-axis. Among the 801 X-ray sources not identified with the \citeauthor{MorauxArtemenko2013} catalog, we found h~Per member counterparts for other 251 X-ray sources (with three sources, X source ID = 4, 273, and 858, having a double identification), of which nine are expected spurious identifications. The expected number of spurious identification is higher than for the identifications with the \citeauthor{MorauxArtemenko2013} sources because of the higher source density in the \citeauthor{CurrieHernandez2010} catalog.

Our final X-ray selected catalog of h~Per members, listed in Table~\ref{tab:xmemb}, is composed of 452 X-ray detected h~Per members, 201 of which have a measured rotational period, and 420 have apparent magnitudes $m_{\rm V}$ and $m_{\rm I_{C}}$. In the second part of Table~\ref{tab:xmemb} we also report the list of the 339 h~Per members with measured rotational period that are not detected in X-rays.


\subsection{Parameter determination}
\label{pardet}

Our aim is to study the relationship between activity and rotation in h~Per members. The stellar activity level is well probed by the X-ray luminosity $L_{\rm X}$, and in particular by the ratio between $L_{\rm X}$ and the bolometric luminosities $L_{\rm bol}$. The dynamo efficiency is expected to depend on the Rossby number $Ro$, that is, on the ratio between the stellar rotational period $P_{\rm rot}$ and the convective turnover time $\tau$. The available measurements from which we started are the X-ray fluxes obtained from the {\it Chandra} data and the rotational period and mass are taken from the \citeauthor{MorauxArtemenko2013} catalog, and from the \citeauthor{CurrieHernandez2010} catalog we retrieved the $m_{\rm V}$ and $m_{\rm I_{\rm C}}$ apparent magnitudes together with (whenever available) spectral type, individual $E(B-V)$, and luminosity class. Below we describe how, starting from these available measurements, we computed $L_{\rm X}$, $L_{\rm bol}$, and $\tau$ together with other parameters needed for their estimation.

\begin{figure}
\centering
\includegraphics[width=8.5cm]{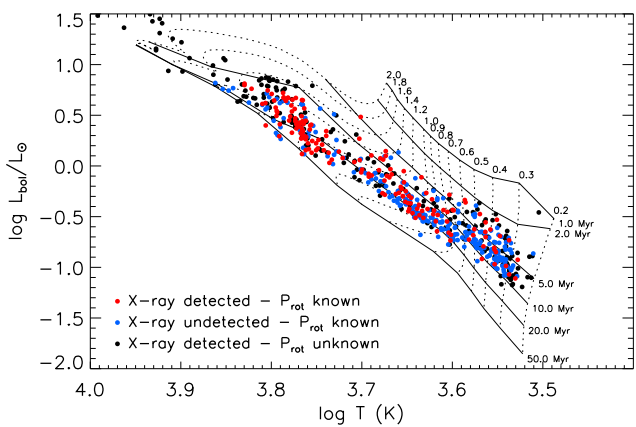}
\caption{HR diagram of h~Per members. Evolutionary models, with labels indicating masses (in solar units) and ages, are taken
from \citet{SiessDufour2000}.}
\label{hrdiagram}
\end{figure}

\subsubsection{Reddening and absolute magnitudes}

To derive the interstellar absorption of each h~Per member, we started from the $E(B-V)$ values published by \citet{CurrieHernandez2010}. Individual $E(B-V)$ values are only available for h~Per members observed with optical spectroscopy and range from $0.188$ to $1.137$, with an uncertainty of 0.013. We adopted their average value, $E(B-V)=0.55\pm0.1$, as reddening for the other h~Per members. From the $E(B-V)$ we derived the $A_{\rm V}$ and $A_{\rm I}$, \citep[assuming $A_{\rm V}=3.12\times E(B-V)$ and $A_{\rm I}=1.87\times E(B-V)$, as reported in ][]{CurrieHernandez2010}, and then inferred the hydrogen column density $N_{\rm H}$, adopting an $N_{\rm H}$ to $A_{\rm V}$ ratio of $2.21\times10^{21}\,{\rm cm^{-2}}$ \citep{GueverOzel2009}. The $A_{\rm V}$ and $A_{\rm I}$ values associated with each source, together with the known cluster distance, allowed us to compute the absolute $M_{\rm V}$ and $M_{\rm I_{\rm C}}$ magnitudes and dereddened $(V-I_{\rm C})$ color for those sources with known apparent $m_{\rm V}$ and $m_{\rm I_{\rm C}}$ magnitude.

\subsubsection{Effective temperature}

For h~Per members observed with optical spectroscopy by \citet{CurrieHernandez2010} and classified as luminosity class~V stars, we derived the effective temperatures $T_{\rm eff}$ by converting the reported spectral class following \citet{KenyonHartmann1995}. For the other h~Per members with no spectroscopic information, but with known absolute $M_{\rm V}$ and $M_{\rm I_{\rm C}}$ magnitude, we derived the effective temperatures $T_{\rm eff}$ from their intrinsic $(V-I_{\rm C})$ color, according again to \citet{KenyonHartmann1995}. The errors on $T_{\rm eff}$ were obtained by propagating the errors on spectral class or $(V-I_{\rm C})$; the obtained uncertainties are smaller than 10\% for $T<10^4$\,K, and increase to 30-50\% for hotter stars.

\subsubsection{Bolometric luminosity}

We derived the stellar bolometric luminosity $L_{\rm bol}$ starting from the absolute $M_{\rm V}$ magnitude and converting it into $L_{\rm bol}$ by adopting the \citet{KenyonHartmann1995} bolometric corrections corresponding to the evaluated $T_{\rm eff}$ of each source. We computed the uncertainties on $L_{\rm bol}$ taking into account both the uncertainty on $M_{\rm V}$ and bolometric corrections. From the inferred $L_{\rm bol}$ and $T_{\rm eff}$ values we constructed the HR diagram of the h~Per members considered, as shown in Fig.~\ref{hrdiagram}.

\subsubsection{X-ray luminosity}

We derived the X-ray luminosity of h~Per members by multiplying their observed X-ray photon flux by $k\,4{\rm \pi}\,d^2$, with $d$ being the cluster distance and $k$ the average photon energy. The observed X-ray fluxes provided by the source detection algorithm correspond to the whole exposure time. Therefore the X-ray luminosities obtained correspond to values averaged over a time interval of $190$\,ks. This interval is long enough to ensure that individual strong flares do not affect the spread in the activity-rotation pattern \citep{PreibischKim2005}. The parameter $k$ depends on the adopted energy band, on the shape of the emitted spectrum, and on the interstellar absorption. We considered the $0.5-8.0$\,keV energy band\footnote{Previous studies on activity-rotation relation where based on the {\it ROSAT} $0.1-2.4$ keV band and not on the {\it Chandra} $0.5-8.0$\,keV band. This difference implies that a correction must be considered in the saturated $L_{\rm X}/L_{\rm bol}$ level observed for MS stars to compare it with our data. \cite{WrightDrake2011} found a saturated level of $\log L_{\rm X}/L_{\rm bol}=-3.13$ and computed a factor of 0.676 to convert $L_{\rm X}$ from the $0.5-8.0$\,keV to $0.1-2.4$\,keV band. This means that the saturation level should be set to $\log L_{\rm X}/L_{\rm bol}=-2.96$. For simplicity we set it to $\log L_{\rm X}/L_{\rm bol}=-3$, since in the following analysis this value is used only as a reference.} and assumed that the X-rays are emitted by an optically thin plasma, at a temperature of $10$\,MK, with heavy element abundances of $0.2$ in solar units \citep[this temperature and metallicity are typical values for low-mass PMS stars of a few Myr,][]{ArgiroffiFavata2006}. Taking into account the hydrogen column density of each source, we obtained $k$ values ranging between $2.5\times10^{-9}$ and $1.1\times10^{-8}\,{\rm erg\,ph^{-1}}$. The derived X-ray luminosities of h~Per members are reported in Table~\ref{tab:xmemb}.

\subsubsection{Convective turnover time}
\label{taumethod}

To compute the Rossby number $Ro=P_{\rm rot}/\tau$, it is necessary to estimate the convective turnover time $\tau$ for each star. While the rotational period is directly measured, $\tau$ is not directly observable. Two different convective turnover times are usually defined: the {\it \textup{global}} convective turnover time, $\tau_g$, defined as the time taken by the plasma to rise through the convective envelope, and the {\it \textup{local}} convective turnover time, $\tau_l$, defined considering not the whole convective envelope, but only a characteristic length at its base. Previous studies usually adopted the local convective turnover time $\tau_l$ to compute the Rossby number because the $\alpha\Omega$ type dynamo is expected to originate at the base of the convective envelope. For MS stars $\tau_l$ is commonly estimated empirically starting from the $(B-V)$ color \citep[e.g.,][]{NoyesHartmann1984,PizzolatoMaggio2003}, a proxy of stellar temperature. h~Per is an intermediate-age PMS cluster, many of the h~Per members lie well above the main sequence (see Fig.~\ref{hrdiagram}), indicating that they are still contracting, and that their internal structure is not settle to the MS status. The $\tau$ of PMS stars can only be estimated by referring to stellar evolutionary models that provide details on the stellar internal structure \citep[e.g.,][]{FeigelsonGaffney2003,PreibischKim2005,AlexanderPreibisch2012}. As a consequence, results based on the inferred $\tau$ values might depend on the accuracy of stellar evolutionary models\footnote{It is known that PMS evolutionary models show systematic mass-dependent discrepancies with observations \citep{HillenbrandBauermeister2008}. These discrepancies also emerge for h~Per, where the comparison between empirical and theoretical cluster sequences suggests a younger age for low-mass members than for higher mass members (Fig.~\ref{hrdiagram}).}.

\begin{figure}
\centering
\includegraphics[width=8.5cm]{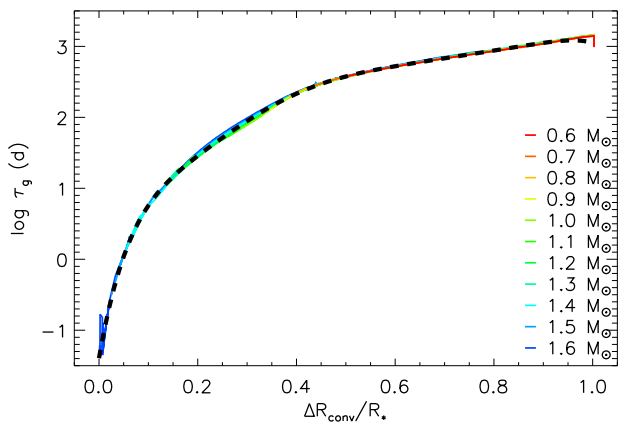}
\caption{Global convective turnover time, $\tau_g$, vs relative convective envelope radius, $\Delta R_{\rm conv}/R_{\star}$, derived from the PMS  evolutionary models of \citet{VenturaZeppieri1998}, with ages ranging between 1 and 30\,Myr. The black dashed line indicates the best-fit function adopted to convert the inferred convective radius of h~Per members to derive their $\tau_g$.}
\label{tauvsrconv}
\end{figure}

To derive the convective turnover time we followed the same procedure as was used by \citet{FlaccomioMicela2004} for PMS stars in the Orion nebula cluster. This procedure is based on PMS evolutionary models of \citet[][and private communication]{VenturaZeppieri1998} and allows estimating the global convective turnover time $\tau_g$. These models provide $\tau_g$ values that depend almost only on the relative dimension of the convective envelope with respect to the stellar radius, $\Delta R_{\rm conv}/R_{\star}$. To show that we report in Fig.~\ref{tauvsrconv} the tracks followed by stars of different mass ($0.6\,{\rm M_{\sun}}\le M \le1.6\,{\rm M_{\sun}}$) during their evolution toward the MS ($1\,{\rm Myr}\le$ {\rm age} $\le30\,{\rm Myr}$) in the $\tau_g$ vs $\Delta R_{\rm conv}/R_{\star}$ space: these tracks perfectly overlap and allow identifying one well-defined relation between $\tau_g$ and $\Delta R_{\rm conv}/R_{\star}$, regardless of stellar mass and age. As a first step, we therefore derived the ratio $\Delta R_{\rm conv}/R_{\star}$ for each star by comparing its position on the HR diagram with the evolutionary tracks of \citet{SiessDufour2000}, that provide this information. We also computed the uncertainty on $\Delta R_{\rm conv}/R_{\star}$ considering the uncertainties on $T_{\rm eff}$ and $L_{\rm bol}$. Then, we inferred the $\tau_g$ corresponding to the estimated $\Delta R_{\rm conv}/R_{\star}$ values, adopting the best-fit relation obtained from the \citeauthor{VenturaZeppieri1998} models\footnote{We note that the use of two different evolutionary models, i.e. \citet{VenturaZeppieri1998} and \citet{SiessDufour2000}, is required because the \citeauthor{VenturaZeppieri1998} models that provide $\tau_g$ consider only stars with masses ranging between 0.6 and $1.6\,{\rm M_{\sun}}$, while the \citeauthor{SiessDufour2000} models also allow exploring stars down to $0.1\,{\rm M_{\sun}}$. We therefore assume that the $\Delta R_{\rm conv}/R_{\star}$ vs $\tau_g$ relation of \citet{VenturaZeppieri1998} also holds for lower mass stars.} that are shown in Fig.~\ref{tauvsrconv}. This method, being based on stellar position on the HR diagram, has the advantage of taking into account possible age spreads. The stellar formation process appears to be not instantaneous, but to last a few Myr \citep[e.g.,][]{BurninghamNaylor2005}. This might produce significant differences in the internal structures of stars with similar mass at the
age of h Per.

\begin{figure}
\centering
\includegraphics[width=8.5cm]{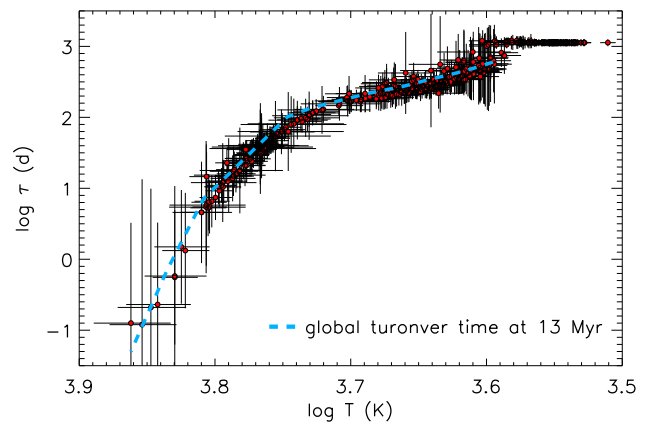}
\caption{Red circles indicate the global convective turnover time inferred for h~Per members. The blue dashed line shows the global convective turnover time provided by \citet{VenturaZeppieri1998} for 13\,Myr old stars.}
\label{tauvstemp}
\end{figure}

In Fig.~\ref{tauvstemp} we show the $\tau_g$ values vs $T_{\rm eff}$ obtained for h~Per members with known rotational period (either detected in X-rays or not), reporting also the resulting error bars on $T_{\rm eff}$ and $\tau_g$. As shown in the plot, the estimated values scatter around the predicted value for 13\,Myr old stars (blue dashed line), as expected. The coolest stars in our sample ($\log T_{\rm eff} \lesssim 3.6$) have $\tau_g$ values that saturate at $\sim1200$\,d. These stars appear slightly younger than the estimated h~Per age because of their position in the HR diagram, and their inferred internal structure is that of fully convective stars (i.e., $\Delta R_{\rm conv}/R_{\star}=1$).

As specified above, with this procedure we inferred the global convective turnover time, $\tau_g$. We decided to use the $\tau_g$ values in our analysis instead of $\tau_l$ because $\tau_g$ can also be defined for fully convective stars, other than for stars with outer convective envelope ($\tau_l$ can instead be defined only for the latter). This choice allows us to include both these stellar categories in our analysis, and investigate whether their activity behaves differently. However, we note that the $\tau_g$ and $\tau_l$ values provided by the \citet{VenturaZeppieri1998} models differ by a factor $\sim3$ regardless of the stellar temperature. Hence the choice of using $\tau_g$ or $\tau_l$ in this work does not affect any possible activity-rotation observed trend, but can only change the absolute $Ro$ values at which trends or regimes occur.


\section{Properties of the X-ray selected catalog of h Per members}
\label{sampleprop}


Our X-ray survey of h~Per cluster and its comparison with the catalogs of \citet{MorauxArtemenko2013} and \citet{CurrieHernandez2010} provided a list of 452 X-ray sources identified as h~Per members. Of these, 201 have a measured rotational period. We also computed upper limits in the X-ray band to other 339 h~Per members with known rotational periods (see Table~\ref{tab:xmemb}). The HR diagram of the h~Per members considered is shown in Fig.~\ref{hrdiagram}.


We based our activity-rotation analysis on the subsample h~Per members with known rotational period and effective temperature, both X-ray detected and undetected. This subsample is composed of 414 stars with masses ranging between 0.32 and 1.6\,${\rm M_{\odot}}$, periods ranging between 0.15 and 16\,d, and Rossby numbers ranging between $2.2\times10^{-4}$ and 6.4. Of these 414 stars, 169 were detected in the X-ray {\it Chandra} observation, showing $L_{\rm X}$ ranging between $3.5\times10^{29}$ and $1.1\times10^{31}\,{\rm erg\,s^{-1}}$. In the whole subsample of 414 stars, 105 where flagged as candidate binaries by \citet{MorauxArtemenko2013}. However, in the analysis presented in this work, we do not distinguish between single or binary systems because the results obtained were the same for the two stellar subsets, which supports the finding of \citet{WrightDrake2011}, who did not find any significant difference in the activity-rotation behavior of single vs binary stars.


The contamination of field stars in the h~Per period catalog from \citet{MorauxArtemenko2013} is expected to be $\sim2\%$. Because our catalog is X-ray selected, it is conceivable that the final contamination of field stars is even lower. In parallel we estimated the spurious identifications of X-ray sources to be $\sim2\%$. Therefore the whole contamination of erroneous sources (because of incorrect identification or false membership) in our subsample is expected to be negligible.

\begin{figure}
\centering
\includegraphics[width=8.5cm]{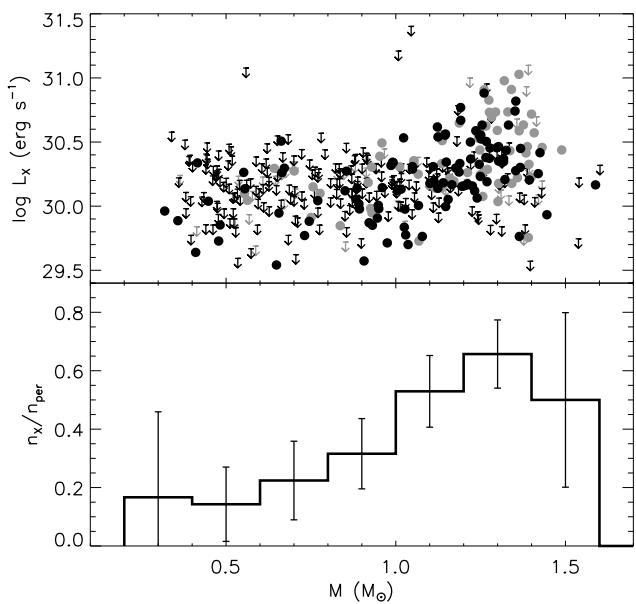}
\caption{{\it Upper panel}: $L_{X}$ vs mass for h~Per members, with black and gray symbols indicating single and binary stars, respectively. {\it Lower panel}: Fraction of sources of the \citet{MorauxArtemenko2013} catalog detected in the X-ray band.}
\label{figlxvsmass}
\end{figure}

The period catalog of \citeauthor{MorauxArtemenko2013} is characterized by a period detection rate that ranges between $\sim80$\% for stars at $I_{CFHT}=15.5$ (corresponding to $M\sim1.5\,{M_{\sun}}$), and $\sim5$\% at $I_{CFHT}=19.5$ (corresponding to $M\sim0.3-0.4\,{M_{\sun}}$). Our X-ray survey, being a flux-limited survey, is likewise characterized by a detection rate that decreases for lower mass stars. In the upper panel of Fig.~\ref{figlxvsmass} we show the mass vs X-ray luminosity scatter plot, including the upper limits of X-ray undetected h~Per members as well, and distinguishing between single and binary stars. In the lower panel of Fig.~\ref{figlxvsmass} we report the detection rate in the X-ray band with respect to the \citeauthor{MorauxArtemenko2013} catalog: the fraction of stars detected in the X-ray band is $\sim15$\% for $\sim0.5\,M_{\odot}$ stars and increases to $\sim70$\% for $\sim1.3\,M_{\odot}$. Therefore the completeness of X-ray detected h~Per subsample, with respect to the total cluster population, is $\sim40\%-50$\% for stars with $M\sim1.3-1.5\,{M_{\sun}}$, and lower than 1\% for stars with $M\sim0.3-0.4\,{M_{\sun}}$. We note moreover that $P_{\rm rot}$ and $L_{\rm X}$ detections are highly correlated, since the most active stars on average display both higher X-ray emission and larger amplitude in their photometric variability, which increases the period detection rate. Therefore the X-ray detected h~Per members our sample, for each mass bin, are preferentially the most active stars.

\begin{figure}
\centering
\includegraphics[width=8.5cm]{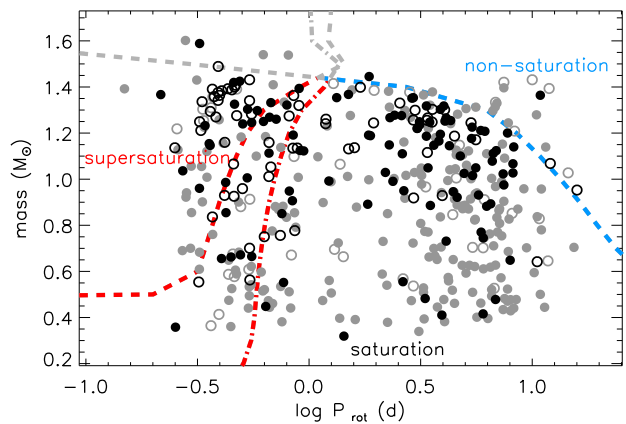}
\caption{Mass vs period of X-ray selected h~Per members detected (black) and undetected (gray) in X-rays. Filled and open symbols indicate single and binary stars, respectively. Red lines separate the loci corresponding to supersaturation and saturation, assuming centrifugal stripping (dash-dotted line) or polar updraft (dashed line) theories. The blue dashed line marks the expected separation between non-saturation and saturation.}
\label{figmassvsprot}
\end{figure}

We first checked which activity regimes can be explored with the 13\,Myr old stellar sample obtained. To this aim we show in Fig.~\ref{figmassvsprot} the scatter plot of mass vs period of h~Per members, including detected and non-detected X-ray sources.

For MS stars the separation between saturated and non-saturated regimes occurs at fixed $Ro$ values, regardless of stellar mass. When $Ro$ is computed as $P_{\rm rot}/\tau_l$, this threshold occurs at $Ro\approx0.13$ \citep{WrightDrake2011}, which turns into $Ro\approx0.04$ when it is taken into account that $\tau_g\approx3\tau_l$. Therefore, assuming that the same applies to intermediate-age PMS stars and considering the internal structures of 13\,Myr old stars of different masses, it is possible to compute rotational periods corresponding to this $Ro$ value. We report in Fig.~\ref{figmassvsprot} this expected separation between non-saturated and saturated regimes (blue dashed line).

We also computed where the supersaturated regime is expected to occur. However, the paucity of observational constraints of supersaturation makes this prediction more uncertain. Hence we considered the two physical mechanisms suggested to explain the supersaturation: centrifugal stripping \citep{JardineUnruh1999} and polar updraft migration \citep{StepienSchmitt2001}. The centrifugal stripping mechanism predicts that coronal emission might be reduced when rotational velocity increases because the largest coronal structures might be disrupted by the centrifugal force. This mechanism takes place when the largest coronal structures extend beyond the co-rotation radius\footnote{The co-rotation radius is defined as the distance from the star at which the orbital period equals the stellar rotational period. Beyond this distance it is not possible to have gravitationally bound structures that rotate with the same angular velocity as the central star.}. Assuming therefore that centrifugal stripping becomes effective when the co-rotation radius is smaller than $3\,R_{\star}$ (i.e., assuming that coronal structures extending up to $2\,R_{\star}$ above the photosphere), as suggested by \citet{WrightDrake2011}, we obtained the dash-dotted red line shown in Fig.~\ref{figmassvsprot}. The polar updraft effect instead predicts that when the rotational velocity increases, then the magnetic flux emergence becomes more efficient near stellar poles, leaving the equatorial regions free of magnetic flux tubes and hence without coronal structures. \citet{WrightDrake2011} quantified the efficiency of this mechanism by defining the parameter

\[
G_{X}=\frac{\omega_{\rm core}^2 R_{\rm core}^3 \sin^2 \theta}{GM_{\rm core}},
\]

\noindent
where $M_{\rm core}$ and $R_{\rm core}$ are the mass and radius of the radiative core, $\omega_{\rm core}$ is its rotational angular velocity, and $\theta$ is the colatitude on stellar surface. Considering MS stars, they suggested that supersaturation occurs where $G_{X}<0.01$. Considering again the internal structures of 13\,Myr old stars, assuming that the core rotates with the same period as the convective envelope and averaging $\sin^2 \theta$ over the stellar surface, we computed the locus corresponding to $G_{X}=0.01$ and report this threshold with a red dashed line in Fig.~\ref{figmassvsprot}. A more detailed discussion on centrifugal stripping and poleward migration is reported in Sect.~\ref{disc}.

The two thresholds that identify the supersaturation regime differ among themselves and with respect to MS stars, where they almost overlap \cite[Fig.~6 in][]{WrightDrake2011}. In particular, the centrifugal stripping threshold for PMS stars is located at periods longer than those of MS stars (because of the larger stellar radii of PMS stars) and longer than those corresponding to the polar updraft threshold. Moreover, the two transitions show large differences for very low mass stars: the polar updraft can operate only in stars with a radiative inner core, and at an age of 13\,Myr, stars with $M\sim0.3-0.5\,M_{\odot}$ have not developed it as
yet.

The thresholds shown in Fig.~\ref{figmassvsprot} indicate that our stellar sample provides a good coverage of supersaturation and saturation regimes. This should allow us to check whether or not at 13\,Myr old these activity regimes are analogous to those of MS stars, or whether, even at this age, PMS stars still show activity levels scattered over several order of magnitudes. In particular, the large number of stars with short rotational periods provide the opportunity of deeply investigating the supersaturated regime, whose occurrence, parameter dependence, and physical origin are still elusive.

\section{Activity vs rotation analysis}
\label{activityanalysis}

We based our activity-rotation analysis on the sample of h~Per members with known rotational period and effective temperature, both X-ray detected and undetected. This sample is composed of 414 stars with masses ranging between 0.3 and 1.6\,${\rm M_{\odot}}$. To check whether the mass is an important parameter in determining the stellar activity level, we divided our stellar sample into different mass bins. The adopted mass grid is $0.3-0.7$, $0.7-1.0$, $1.0-1.2$, $1.2-1.4$, and $1.4-1.6\,M_{\odot}$. The aim is to obtain a minimum number of stars in each mass
bin to perform a statistically meaningful analysis. The last bin, $1.4-1.6\,M_{\odot}$, is poorly populated, but we decided to keep it separated because in this mass range the expected behavior is significantly different from that of lower mass stars (see Fig.~\ref{figmassvsprot}). All but one of the stars populating the first bin, that is, $0.3\,M_{\odot}<M_{\star}<0.7\,M_{\odot}$, are fully convective stars ($\Delta R_{\rm conv}/R_{\star}>0.95$), but some fully convective stars also populate ($\sim30\%$) the second mass bin. Each mass bin is also composed of a significant percentage (ranging between 20\% and 50\%) of candidate binary stars. In our analysis we investigated whether and how different parameters could be relevant for determining stellar activity in PMS stars, and to this aim we considered separately single vs binary stars, and fully convective from partially convective stars. The different subsamples obtained  showed no significant differences, however. Therefore in the analysis presented here we did not consider these classes separately.

\begin{figure*}
\centering
\includegraphics[width=8.5cm]{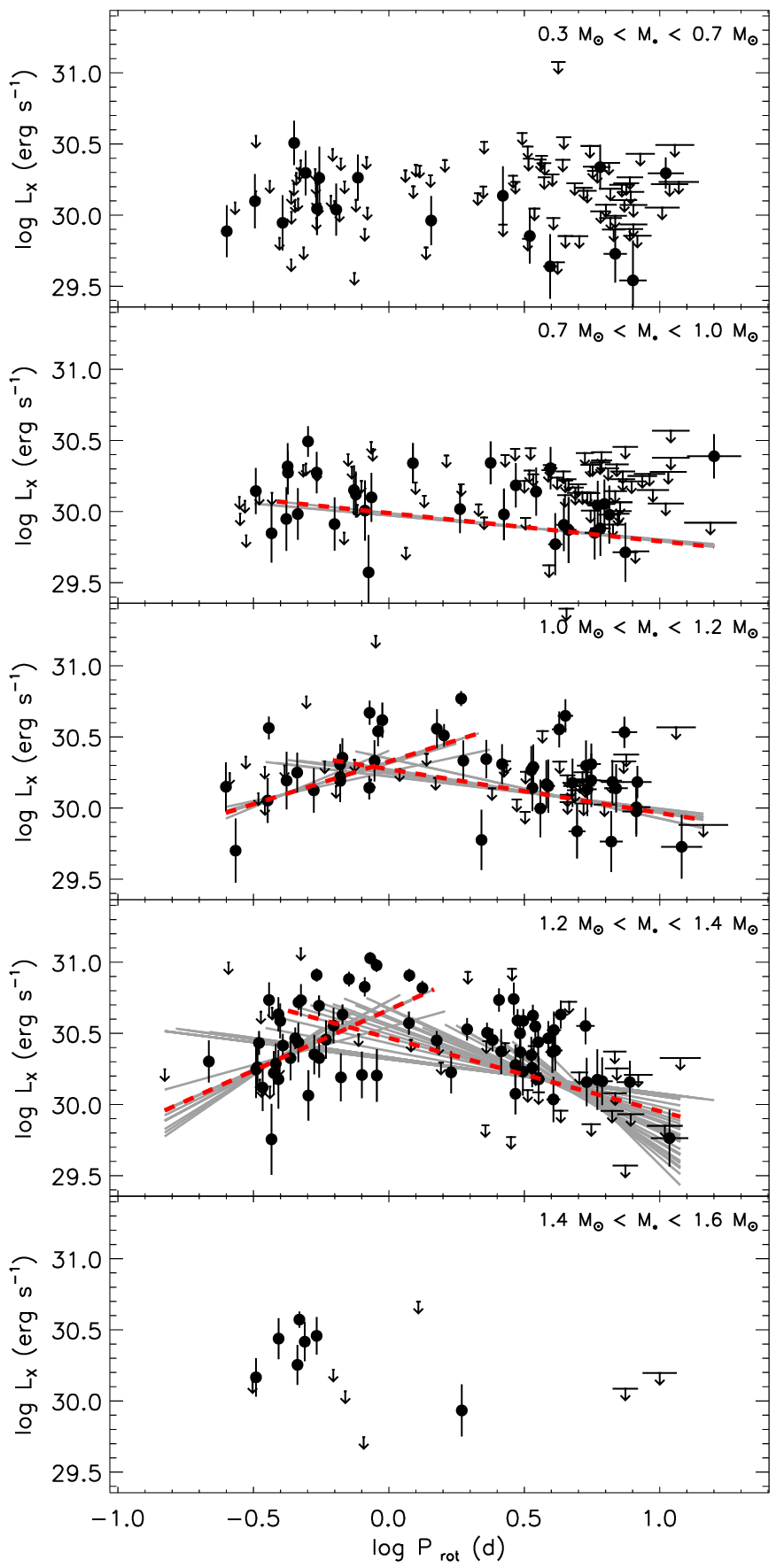}
\includegraphics[width=8.5cm]{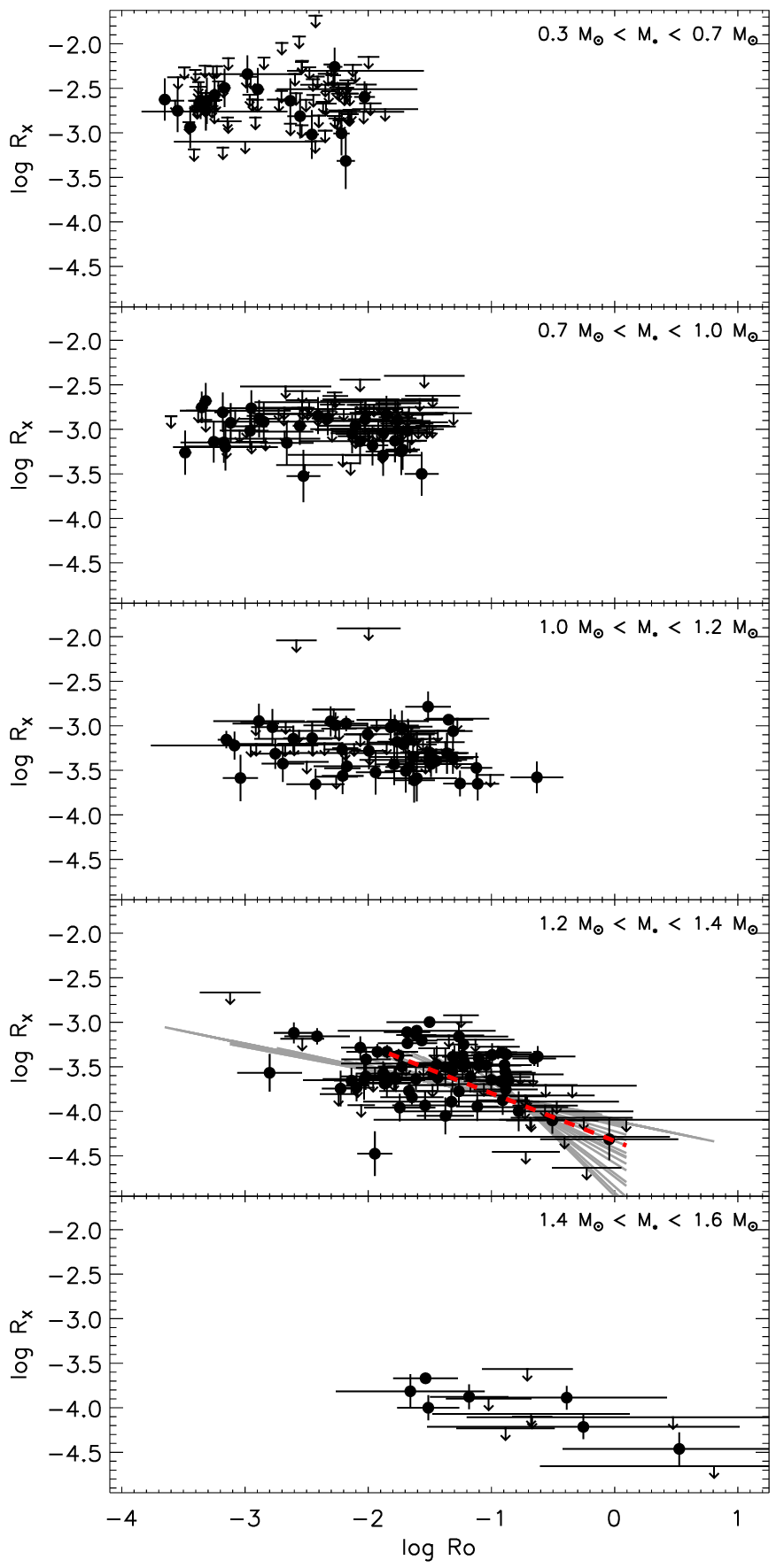}
\caption{{\it Left panel}: X-ray luminosity vs rotational period for h~Per members of different mass. {\it Right panel}: Fractional X-ray luminosity vs Rossby number for h~Per members of different mass. In both cases gray lines mark the best fit corresponding to intervals in which the two quantities have a significant correlation (confidence level higher than 99\%), with red dot-dashed lines indicating the highest significant cases for both positive and negative correlation.}
\label{figlxvsrot}
\end{figure*}

To search for intrinsic correlations among variables, we used the {\tt ASURV} software package \citep{IsobeFeigelson1990,LavalleyIsobe1992}, which implements methods presented in \citet{IsobeFeigelson1986} that provide correlation and regression techniques for samples containing both detections and upper limits. In particular, we searched for significant correlations by using Spearman's rho test and computed linear regression with the EM algorithm.

We searched for a significant correlation between $\log L_{\rm X}$ vs $\log P_{\rm rot}$ considering separately stars with different mass. We separated stars by mass because the completeness of our sample depends on stellar mass, and because we cannot exclude that stellar activity at this age are mass dependent. The resulting scatter plots are shown in the left panels of Fig.~\ref{figlxvsrot}.

Inspecting these plots and considering that MS stars show different activity regimes, we guessed that different correlations corresponding to different regimes may exist also for intermediate-age PMS stars. We did not assume a priori at which $P_{\rm rot}$ value regime switches take place. We therefore investigated separately different $P_{\rm rot}$ subintervals. In particular, for each mass bin, we divided the entire $P_{\rm rot}$ range into two adjacent subintervals, searched for correlations separately in the two subintervals, and repeated this analysis considering different $P_{\rm rot}$ separating values.

We assumed 99\% as confidence level. The best fits corresponding to all the significant correlations obtained are indicated in Fig.~\ref{figlxvsrot} with gray lines. Since in several cases significant correlations are present for a vast range of $P_{\rm rot}$ separating values, we identified the positive and negative correlations corresponding to the highest significance to synthesize these results in a more
compact way. We highlight these correlations in Fig.~\ref{figlxvsrot} with red dashed lines. For some mass bins, none, or only one kind of correlation emerges. For the $1.0-1.2$ and $1.2-1.4\,{M_{\odot}}$ bins we instead found both positive and negative correlations. In these cases we note that positive and negative correlations are present also in non-overlapping and hence independent ranges. This renders the detection of different regimes
statistically reliable. The fact that significant correlations are obtained for a vast range of $P_{\rm rot}$ separating values could pinpoint the uncertainty on the exact $P_{\rm rot}$ value at which these regimes switch, but it could also indicate that the regime switch is not as sharp.

To summarize, and taking into account the highest significance correlations, we found a clear evidence of correlations between $\log L_{\rm X}$ and $\log P_{\rm rot}$ for stars with masses in the ranges $0.7-1.0\,M_{\odot}$, $1.0-1.2\,M_{\odot}$, and $1.2-1.4\,M_{\odot}$. In particular, for stars with masses between $0.7$ and $1.0\,M_{\odot}$, the data indicate a negative correlation over the whole $P_{\rm rot}$ range, with a slope of $-0.20\pm0.07$. For the $1.0-1.2\,M_{\odot}$ and $1.2-1.4\,M_{\odot}$ mass bins we found that for periods shorter than $\sim1$\,d there is a positive correlation corresponding to slopes of $0.60\pm0.16$ and $0.86\pm0.22$. We also observed a significant negative correlation for periods longer than $\sim1$\,d, with slopes of $-0.31\pm0.09$ $-0.51\pm0.09$, respectively.

We also searched for significant correlations between the logarithmic fractional X-ray luminosity, $\log R_{\rm X} = \log L_{\rm X}/L_{\rm bol}$, and $\log Ro$. For MS stars the relation between $R_{\rm X}$ and $Ro$ is the same for stars of different masses. However, we first also inspected the $\log R_{\rm X}$ and $\log Ro$ patterns separating stars of different mass (right panels of Fig.~\ref{figlxvsrot}) because {\it a}) we wished to check whether in intermediate-age PMS stars activity depends on mass, {\it b}) the completness of our stellar sample depends on stellar mass, and {\it c}) we aim at minimizing the effect of possible mass-dependent systematic uncertainty present in the evolutionary models used to infer stellar $\tau$. After dividing the stars into different mass bins, it might be expected that a significant correlation between $\log L_{\rm X}$ and $\log P_{\rm rot}$ would correspond to a significant correlation between $\log R_{\rm X}$ and $\log Ro$, but this is not the case: in all the inspected cases we did not find evidence of significant correlations, with the only exception of a negative correlation for stars with $1.2\,M_{\odot}<M<1.4\,M_{\odot}$ and $\log Ro \gtrsim -2$, corresponding a slope of $-0.54\pm0.09$.

\begin{figure}
\centering
\includegraphics[width=8.5cm]{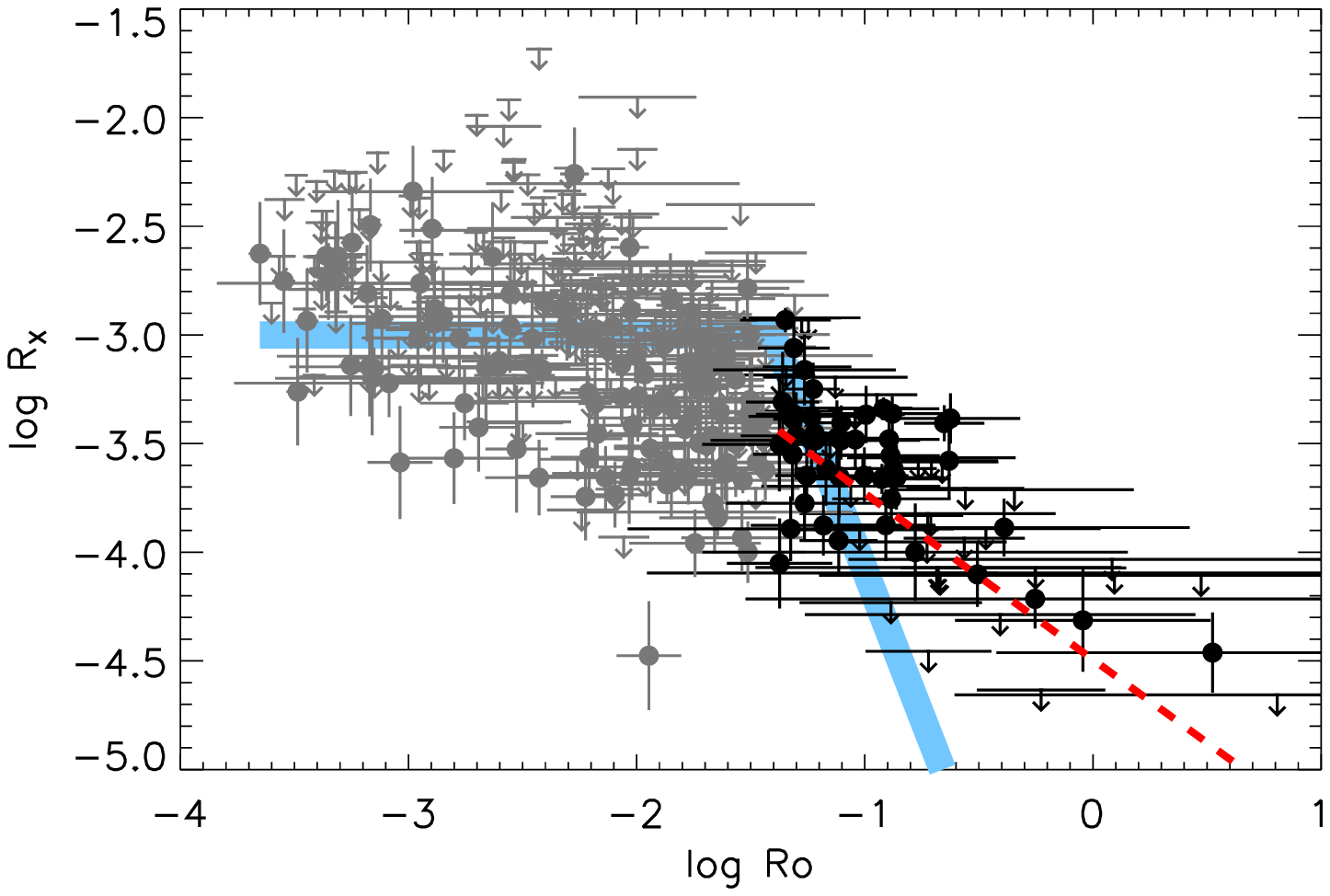}
\caption{Fractional X-ray luminosity vs Rossby number for h~Per members. The light blue solid line indicates the saturated and non-saturated relation obtained for MS stars by \citet{WrightDrake2011}. Black symbols mark h~Per members used to perform the best fit in the expected non-saturation regime (i.e., $Ro>0.04$ and $M>1.0\,M_{\odot}$), with the red dashed line being the best-fit relation. Gray symbols indicate the remaining h~Per members.}
\label{figrxvsroall}
\end{figure}

Inspecting the $\log R_{\rm X}$ vs $\log Ro$ scatter plots of Fig.~\ref{figlxvsrot}, we note that even if significant correlations in most cases do not emerge, the typical $\log R_{\rm X}$ level decreases for increasing stellar mass. We cannot distinguish whether this effect is due to an intrinsic difference in the typical $\log R_{\rm X}$ level of stars of different mass or simply a consequence of the different X-ray sensitivity in different mass bins. We selected our stellar sample from a flux-limited X-ray survey (see Sect.~\ref{sampleprop}), hence the survey reaches the lowest $R_{\rm X}$ for the highest mass stars. If the latter is the case, then lower mass stars might have a larger scatter in their $\log R_{\rm X}$ distribution than higher mass stars.

To investigate the $R_{\rm X}$ vs $Ro$ relation for intermediate-age PMS stars in more detail, we considered the whole stellar sample, regardless of the different stellar masses, in analogy to the $R_{\rm X}$ vs $Ro$ trends observed for MS stars. We show the $\log R_{\rm X}$ vs $\log Ro$ scatter plot of h~Per members of our sample in Fig.~\ref{figrxvsroall}, where we report as a reference the best-fit relation describing saturated and non-saturated regimes that was obtained by \citet{WrightDrake2011} for MS stars, which we adjusted for the different $\tau$ used here. The whole stellar sample indicates a significant correlation over the whole $Ro$ ranges with fractional X-ray luminosity increasing for decreasing Rossby number. As stressed above, this correlation might be due to the different X-ray sensitivity of our sample for different stellar masses: the lowest $R_{\rm X}$ values can be explored only for the highest mass stars, which on average have lower $\tau$ and hence higher $Ro$. This selection effect is evident from the left panel plots of Fig.~\ref{figlxvsrot}. To avoid this mass selection effect, we searched for correlations considering only stars with $M_{\star}>1.0\,M_{\odot}$, since our completeness is uniform over this mass range (see Fig.~\ref{figlxvsmass}). In particular, examining only the $Ro$ range corresponding to non-saturation (i.e., $Ro>0.04$), we found a significant negative correlation, corresponding to a slope of $-0.76\pm0.10$, which
is shown by the red dashed line in Fig.~\ref{figrxvsroall}.

From this analysis we found evidence that for long rotational periods (or large $Ro$) X-ray luminosity (or $R_{\rm X}$) decreases for increasing $P_{\rm rot}$ (or $Ro$). We also found that for short rotational periods, stars with $1.0\,M_{\odot}<M<1.4\,M_{\odot}$ show a positive correlation between $L_{\rm X}$ and $P_{\rm rot}$. This trend is not observed in the $R_{\rm X}$ vs $Ro$ space.

Before discussing these results, and in particular the positive correlation between $L_{\rm X}$ and $P_{\rm rot}$ that is not observed between $R_{\rm X}$ and $Ro$, we performed some additional checks. As a first step, to distinguish whether this positive correlation disappears in the $R_{\rm X}$ vs $Ro$ space because of the $L_{\rm X}$ to $R_{\rm X}$ conversion or because of the
change from $P_{\rm rot}$ to $Ro$ , we inspected the $R_{\rm X}$ vs $P_{\rm rot}$ space. We found a significant positive correlation between $R_{\rm X}$ vs $P_{\rm rot}$  for the same mass bins and for short periods. This indicates that this positive correlation emerges only when $L_{\rm X}$ or $R_{\rm X}$ are considered with respect to $P_{\rm rot}$ and that it is not related to the known positive correlation between $L_{\rm X}$ and $L_{\rm bol}$ or $L_{\rm X}$ and stellar mass.

As already stressed in Sect.~\ref{taumethod}, the $\tau$ estimation is particularly critical for PMS stars because empirical relations calibrated on MS stars cannot be used. For PMS stars $\tau$ must be estimated considering stellar evolutionary models and inferring the internal structure of each star. Therefore we scrutinized the method adopted for estimating $\tau$ to try to evaluate whether and how our results depend on the peculiar evolutionary models considered, or on the method adopted.

We checked the theoretical $\tau$ values of \citeauthor{VenturaZeppieri1998}, comparing them with the value provided by \citeauthor{LandinMendes2010}. The two models provide very similar values both for $\tau_l$ and $\tau_g$. The \citeauthor{LandinMendes2010} models, however, range only from 0.6 to $1.2\,M_{\odot}$, which is at odds with the 0.6 to $1.6\,M_{\odot}$ range covered by \citeauthor{VenturaZeppieri1998} The $\tau$ vs $M$ relation at 13 \,Myr changes significantly in slope in the $1.0\,M_{\odot}-1.6\,M_{\odot}$ range (this slope change also emerges in the $\tau$ vs $T_{\rm eff}$ relation at $\log T_{\rm eff} \approx 3.75$, see Fig.~\ref{tauvstemp}). This is because of the steep $\tau$ decrease during final PMS contraction stages, just before arriving at the zero-age main sequence. Consequently, considering the mass range of our stellar sample, we preferred to use the \citeauthor{VenturaZeppieri1998} models to derive $\tau$ , since they provide a better coverage of this mass range.

We examined whether and how the method employed to estimate $\tau$ , based on the position of each star in the HR diagram, affects the results. We performed different $\tau$ estimations, using a 13\,Myr isochrone, and inferring $\tau$ from the $T_{\rm eff}$ or $L_{\rm bol}$ only. Both these estimates  rely on assuming that all the stars are exactly coeval, and basing the $\tau$ estimation on only one stellar parameter. With these different $\tau$ derivation techniques the $Ro$ vs $R_{\rm X}$ correlation search provides exactly the same results: a significant negative correlation for large $Ro$ (the non-saturation regime), and no significant correlation for low $Ro$.

This comparison between models and methods demonstrates that they are all similar and that a specific choice of model and/or method does not significantly affect the inferred $\tau$. Available evolutionary models are known to have mass-dependent systematic uncertainties \citep{HillenbrandBauermeister2008}, and hence possibly future model improvements might change our understanding of the $R_{\rm X}$ vs $Ro$ correlation in PMS stars. We are confident that our analysis, performed on different mass bins, minimizes the effect of these possible systematic errors on the obtained results.

We also considered the random uncertainties on $Ro$, and in particular whether these errors, being larger than that on $P_{\rm rot}$, could hide an underlying correlation. We evaluated that even if some stars have very large uncertainties, the average $Ro$ error in the different mass bins is significantly smaller than the $Ro$ range we explored. This suggests that random uncertainties are not large enough to completely bury a correlation, if present.

These checks and considerations indicate that the obtained results, and in particular the absence of any positive correlation between $Ro$ and $R_{\rm X}$ in the low $Ro$ range, which is at odds with that observed between $P_{\rm rot}$ and $L_{\rm X}$, appears to be independent of systematic and/or random uncertainties in the inferred $Ro$. Therefore the positive correlation between activity and rotation occurring for fast rotators with $1.0\,M_{\odot}<M<1.4\,M_{\odot}$, a pattern analogous to the supersaturation phenomenon, appears to be related to the stellar rotational period and not to the Rossby number.

\section{Discussion}
\label{disc}

The stellar sample composed of 414 h~Per members offered us the possibility of investigating correlations between activity and rotation in stars at 13\,Myr. The first result we obtained is that h~Per members, depending on mass, show relations between activity, probed by $L_{\rm X}$, and rotation. In particular, inspecting the $L_{\rm X}$ vs $P_{\rm rot}$ space, we found a positive correlation for $P_{\rm rot}\lesssim1$\,d for masses ranging between $1.0\,M_{\odot}$ and $1.4\,M_{\odot}$, while we found a negative correlation between $L_{\rm X}$ and $P_{\rm rot}$ for long $P_{\rm rot}$ for $0.7\,M_{\odot}<M<1.4\,M_{\odot}$. Conversely, for lower mass stars ($M_{\star}<0.7\,M_{\odot}$) we did not observe correlations between $L_{\rm X}$ and $P_{\rm rot}$, with $L_{\rm X}$ displaying a significant scatter, and being compatible with a constant value, regardless of $P_{\rm rot}$.

In the $R_{\rm X}$ and $Ro$ space no clear relations were observed when we separately considered the different mass bins, with only one exception of a negative correlation in the $1.2-1.4\,M_{\odot}$ bin. This negative correlation also emerges significantly when all the $M_{\star}>1.0\,M_{\odot}$ stars are considered, over the $Ro$ range expected for non-saturated regime. No evidence of positive correlations is present in the $R_{\rm X}$ vs $Ro$, regardless of mass and $Ro$ range. Here we discuss whether and how these observed trends are reconcilable to the supersaturated, saturated, and non-saturated regimes that are observed for MS stars, and what can be inferred, from this comparison, about
the coronal properties of intermediate-age PMS stars.

\begin{figure}
\includegraphics[width=8.5cm]{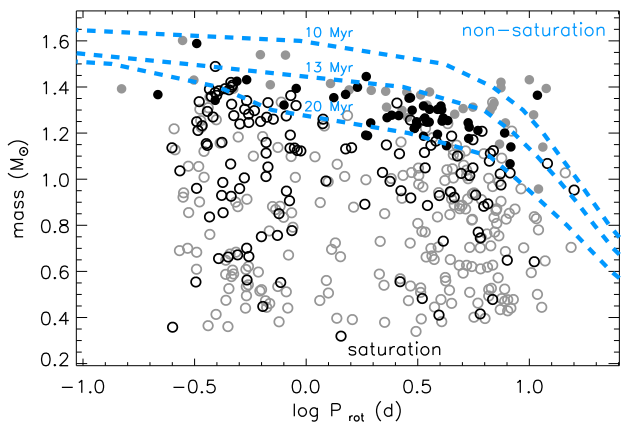}
\caption{Mass vs period of X-ray selected h~Per members detected (black) and non-detected (gray) in X-rays. Filled and open symbols indicate non-saturated and saturated or supersaturated stars. The blue dashed line marks the expected separation between non-saturation and saturation ($Ro=0.04$) at different ages, as labeled in the plot.}
\label{figmassvsprot2}
\end{figure}

\subsection{Saturation and non-saturation}

h~Per stars show clear evidence of a negative correlation between $\log L_{\rm X}$ and $\log P_{\rm rot}$ for stars with $0.7\,M_{\odot}<M_{\star}<1.4\,M_{\odot}$ and for long rotational periods. This negative correlation is contiguous (for $1.0\,M_{\odot}<M_{\star}<1.4\,M_{\odot}$ ) to the positive correlation observed instead for $P_{\rm rot}\lesssim1$\,d.  The fact that these correlations are adjacent indicates that there is no clear evidence of an intermediate behavior between them. We interpret this observed negative correlation as the average effect due to the fact that stars with $1\,{\rm d}\lesssim P_{\rm rot}\lesssim10\,{\rm d}$ are mainly in the saturated regime, with only some of the slowest rotators being located in the non-saturated regime, and hence having a slightly lower $L_{\rm X}$ that provides the global negative correlation. This interpretation is also supported by the inspection of the $R_{\rm X}$ vs $Ro$ space, where we found that stars with $M_{\star}>1.0\,M_{\odot}$ whose $Ro$ correspond to the non-saturated regime indeed display a significant negative correlation.

The fact that we found h~Per members in the non-saturated regime apparently contradicts what was inferred from the mass vs $P_{\rm rot}$ scatter plot and the predicted thresholds between the different regimes (Fig.~\ref{figmassvsprot}). The predicted edge between saturated and non-saturated regimes (the blue dashed line in Fig.~\ref{figmassvsprot}) places almost all the h~Per members in the saturated regime. However, the position of this threshold is uncertain. This threshold is highly sensitive to stellar age: a slightly older age would bring a significant fraction of stars into the non-saturated regime, as shown in Fig.~\ref{figmassvsprot2}. The Rossby number estimate for each h~Per member, obtained taking into account possible age spread, provided several stars with $\log Ro$ larger than the limiting value for saturation ($Ro=0.04$, these stars are indicated with filled circles in Fig.~\ref{figmassvsprot2}). Moreover, it may be inappropriate to assume that the threshold for intermediate-age PMS stars is the same as that for MS stars.

We finally note that all the negative correlations obtained show slopes significantly shallower than that observed for the non-saturated pattern in MS stars \citep[i.e., $\sim-2$ or $\sim-2.7$,][]{PizzolatoMaggio2003,WrightDrake2011}. However, no separation between saturation and non-saturation in our sample is evident, and hence the slope determination is likely affected by partial inclusion of a few saturated stars. Moreover, our survey, being an X-ray limited survey, only allowed us to detect the most active stars in the non-saturated regime. This means that we probe only the upper envelope of the $R_{\rm X}$ vs $Ro$ parameter space, which in turn implies a highly uncertain estimate of the slope.

\subsection{Supersaturation}

h~Per, thanks to its age, is well populated by very fast rotating stars. This makes it a unique benchmark to probe the supersaturated regime. Considering separately stars with different mass, we found clear evidence of a positive correlation between $\log L_{\rm X}$ and $\log P_{\rm rot}$ for stars with $1.0\,M_{\odot}<M_{\star}<1.2\,M_{\odot}$ and $1.2\,M_{\odot}<M_{\star}<1.4\,M_{\odot}$, and for periods shorter than $\sim1$\,d. The slopes in the two cases are $0.60\pm0.16$ and $0.86\pm0.22$, respectively. This pattern is compatible with the supersaturation behavior. Moreover, we observed that the activity level in this regime is better described by the rotational period than by the Rossby number: we did not observe any significant correlation, analogous to the supersaturation, between $R_{\rm X}$ and $Ro$, either by separating stars according to mass or considering the whole sample. As already discussed in Sect.~\ref{taumethod} and \ref{activityanalysis}, we note that this finding is model
dependent, but this result strongly suggests that supersaturation is unrelated to the dynamo efficiency, in contrast to the non-saturated and saturated regimes. Therefore the dynamo-related mechanisms, which consider dynamo feedback effects and suggest that induced magnetic fields could reduce differential rotation and hence the efficiency of the dynamo itself \citep[e.g.,][]{KichatinovRudiger1993,Rempel2006}, are very likely not responsible for supersaturation. The mechanism causing supersaturation has to be searched for among those related to other mechanisms, such as centrifugal stripping or polar updraft migration. Here we discuss our results by comparing them with predictions and indicators that can be deduced from these two mechanisms.

\begin{figure*}
\includegraphics[width=8.5cm]{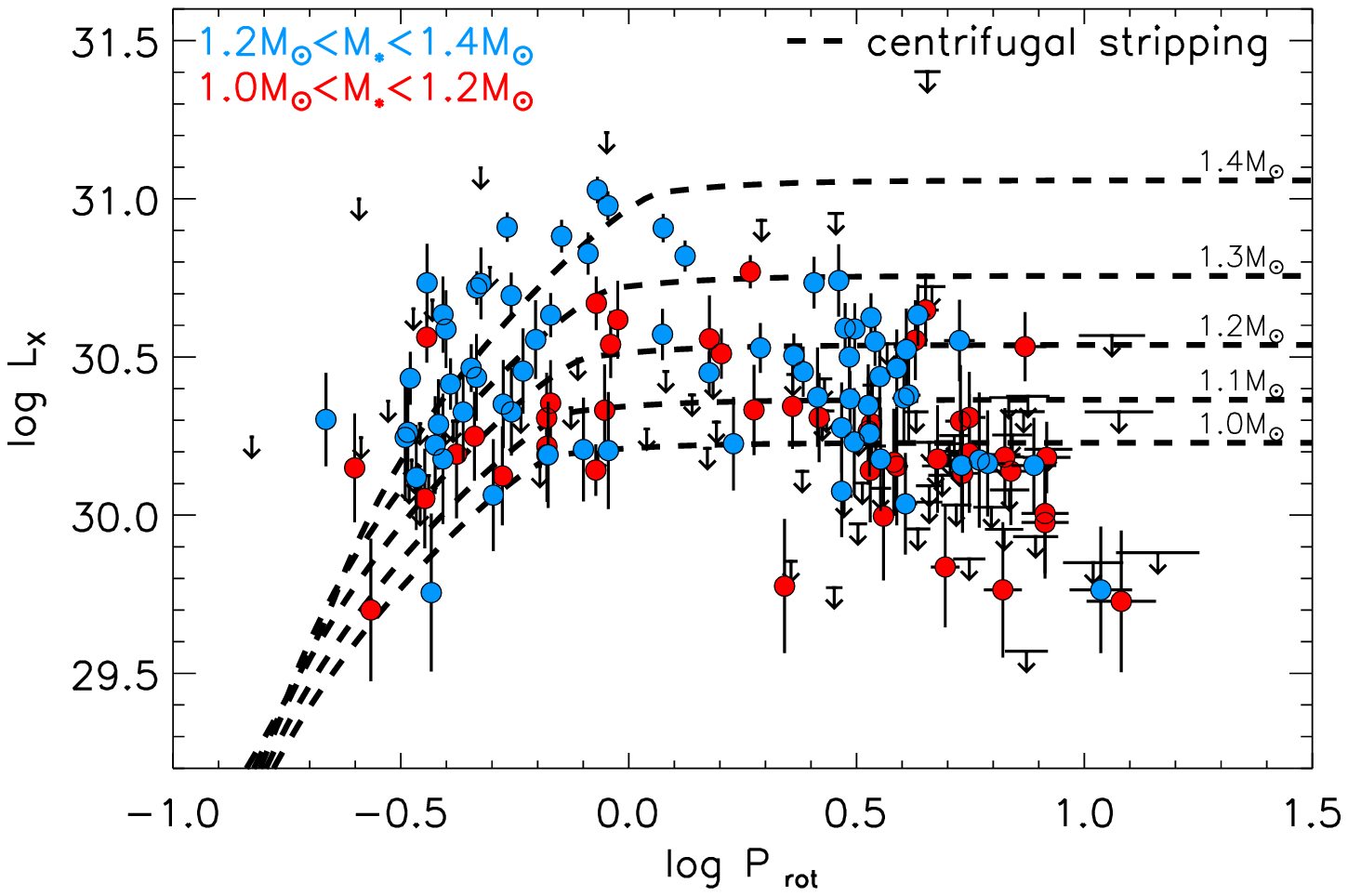}
\hfill
\includegraphics[width=8.5cm]{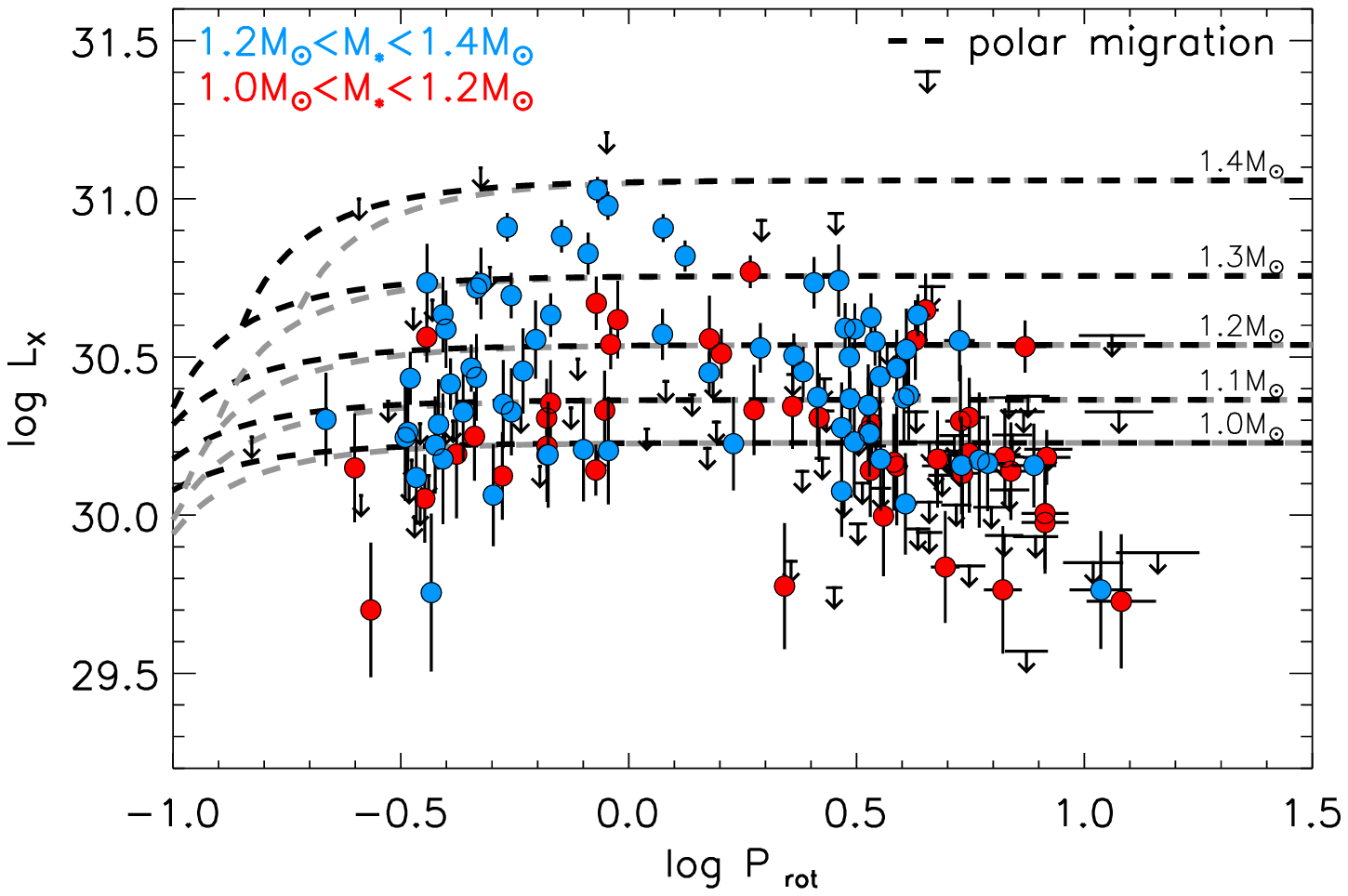}
\caption{X-ray luminosity vs rotational period for h~Per members, with masses ranging between 1.2 and $1.4\,M_{\odot}$ (blue), and 1.0 and $1.2\,M_{\odot}$ (red), compared with supersaturation behavior predicted from centrifugal stripping models ({\it left panel}) or polar updraft migration models ({\it right panel}), obtained considering different stellar masses (dashed gray lines in the {\it right panel} are computed assuming that the radiative core has a 30\% higher rotation than the convective envelope).}
\label{figssmodel}
\end{figure*}

\subsubsection{Centrifugal stripping}

The centrifugal stripping mechanism predicts that in rapidly rotating stars the coronal emission is reduced because the co-rotation radius $R_{\rm cor}$, and hence the volume within the co-rotation radius that is the one available for stable coronal structures, decreases \citep{JardineUnruh1999,Jardine2004}. To try to quantify how, in the centrifugal stripping scenario, $L_{\rm X}$ decreases for increasing rotational velocity, we assumed that $L_{\rm X}$ scales as the available volume and computed the available coronal volume for increasing rotational velocity (i.e., decreasing rotational period). Therefore we set

\[
L_{\rm X} = L_{\rm X0}\frac{V_{\rm red}}{V_{0}}
\]

\noindent
where $V_{0}$ is the coronal volume of a slowly rotating star
that is not affected by centrifugal stripping, $V_{\rm red}$ indicates the reduced coronal volume, and $L_{\rm X0}$ is the X-ray luminosity corresponding to the saturated level. We assumed that the coronal volume of a slowly rotating star is a sphere shell extending from the stellar surface up to $\lambda$ stellar radii above it. \citet{JardineUnruh1999} computed the volume available for the corona, when both magnetic fields and centrifugal forces are taken into account. For an aligned quadrupolar magnetic field\footnote{We considered a quadrupolar magnetic field since young stars usually have complex multipolar fields \citep[e.g.]{HussainJardine2007,GregoryMatt2008,DonatiLandstreet2009}, however, even considering the dipole case provided by \citet{JardineUnruh1999}, the results are very similar.} they found that undisturbed corona extends up to the surface defined by

\[
r = R_{\rm co-rot}\left( \frac{3\cos^2\theta-1}{\sin^2\theta\,(5\cos^2\theta-1)} \right)^{1/3}
\]

\noindent where $\theta$ is the stellar colatitude, and $R_{\rm co-rot}$ is the co-rotation radius,

\[
R_{\rm co-rot}=\left( \frac{GM_{\star}P_{\rm rot}^{2}}{(2\pi)^2} \right)^{1/3}
\]

We computed the available volume, $V_{\rm red}$, at each rotational period, as the intersection of the volume delimited by this surface with the sphere shell describing the coronal volume of a slowly rotating star.

\begin{figure}
\includegraphics[width=8.5cm]{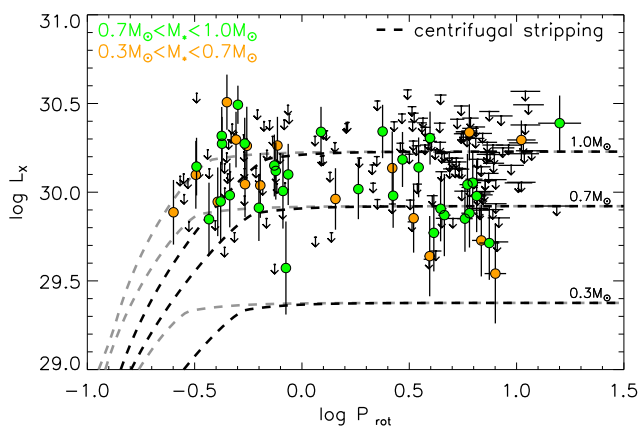}
\caption{X-ray luminosity vs rotational period for h~Per members, with masses ranging between 0.7 and $1.0\,M_{\odot}$ (green), and 0.3 and $0.7\,M_{\odot}$ (orange), compared with supersaturation behavior predicted by centrifugal stripping models (dashed gray and black lines are computed assuming that the coronal extends up to $1\,R_{\star}$ and $2\,R_{\star}$ over the stellar surface).}
\label{figssmodel3}
\end{figure}

The predicted curves obtained are shown in the left panel of Fig.~\ref{figssmodel}, where we set $L_{\rm X0}$ as $10^{-3}$ with respect to the stellar bolometric luminosity, and $V_{0}$ as the volume corresponding to a sphere shell extending from the stellar surface up to two stellar radii above it (i.e., $\lambda=2$). For each stellar mass we assumed for the stellar radius the value predicted by \citet{SiessDufour2000} at an age of 13\,Myr. We
note that the value assumed for $L_{\rm X0}$ determines the vertical position of the predicted curves, while the value of $V_{0}$ determines at which rotational period the coronal stripping starts to occur (i.e., the lower $V_{0}$ the shorter the $P_{\rm rot}$ needed to have centrifugal stripping). Finally, we note that we focused here on saturation and supersaturation patterns, hence any prediction will include only these behaviors, and therefore we compare models and data in the short period range, that is, $P_{\rm rot}\lesssim 1$\,d.

These predicted curves agree quite well with observed data. In particular, the predicted slopes between $\log L_{\rm X}$ and $\log P_{\rm rot}$, which vary between $1.7$ and $1.0$ for $-0.5<\log P_{\rm rot}<0$, are similar to the observed curves. This comparison indicates that despite the simplified assumptions made to compute the predicted $L_{\rm X}$, centrifugal stripping could be the mechanism responsible for the supersaturation. In this scenario the observed threshold of $P_{\rm rot}\sim1$\,d at which the supersaturation occurs, suggests that in $1.0-1.4\,M_{\odot}$ stars at 13\,Myr age, coronal structures extend up to $\sim2\,{R_{\star}}$ above the stellar surface. This value for coronal dimensions agrees with the value inferred by \citet{WrightDrake2011}, which
was obtained by studying supersaturation in G-type and K-type MS stars. Moreover, both the analysis of flaring structures \citep{GetmanFeigelson2008,ArgiroffiFlaccomio2011} and the extrapolation coronal structures from magnetic maps \citep{JohnstoneJardine2014} in PMS stars indicated that coronae can also have structures extending up to several stellar radii, which means that they
are susceptible to coronal stripping.

We did not detect any supersaturation in the low-mass stars included
in our sample, whose X-ray luminosities remained constant down to $P_{\rm rot}\sim0.3$\,d. We show in Fig.~\ref{figssmodel3} the observed $L_{\rm X}$ vs $P_{\rm rot}$ scatter plot for stars with $M<1.0\,M_{\odot}$, compared to predictions based on centrifugal stripping. This comparison indicates that centrifugal stripping was expected to occur for the most rapid rotators of our sample. We stress that in this mass range the completeness of our survey is too low (a few percent) and that we detected only the most active fraction of stars, therefore we cannot state that predictions and observation disagree. However, we could speculate that the dimensions of coronal structures in these stars might be smaller than that of solar-like mass stars. In this case, the $P_{\rm rot}$ threshold for centrifugal stripping should decrease (as shown by the gray dashed lines in Fig.~\ref{figssmodel3}, obtained by assuming coronal dimension of $1\,R_{\star}$). 

With this further assumption all of our data become consistent with predictions based on centrifugal stripping. The hypothesis that centrifugal stripping is the mechanism responsible for supersaturation is also supported by the fact that supersaturation does not seem to be related to a reduced coverage of stellar surface by magnetically active regions, in fact, no evidence of supersaturation has been observed in chromospheric emission \citep{MarsdenCarter2009,ChristianMathioudakis2011}. We finally note that depending on stellar mass and age, the hypothesis that supersaturation is caused by coronal stripping implies that fast rotators of different masses in the $R_{\rm X}$ vs $Ro$ space settle on different tracks, and this can contribute to the observed spread.

\subsubsection{Polar updraft migration}

The alternative mechanism proposed to explain supersaturation, independent of the dynamo efficiency, is the polar updraft migration proposed by \citet{StepienSchmitt2001}. This theory predicts that at high rotational velocity active regions preferentially concentrate near stellar poles, leaving the equatorial region free from magnetic structures and coronal plasma, reducing the filling factor, and hence decreasing the X-ray emission. This polar updraft migration is assumed to be caused by the non-uniform heating that at high rotational velocity characterizes the base of the convective envelope. The emerging energy flux $F$ in the radiative core is proportional to the local effective gravitational acceleration $g_{\rm eff}$ (the von~Zeipel theorem). In rapidly rotating stars, $g_{\rm eff}$ changes significantly with stellar latitude, it is lowest at the equator and highest at the poles. Hence the subsequent non-uniform heating at the base of the convective envelope might favor a magnetic field emergence higher near the stellar poles than in equatorial regions. This polar concentration might also be amplified by the Coriolis force, which tends to deflect the plasma rising across the stellar convective envelope and the frozen magnetic flux tubes toward the stellar poles \citep{SolankiMotamen1997}, even if the effect due to the Coriolis force has been predicted to be negligible with respect to the non-uniform heating at the base of the convective envelope \citep{StepienSchmitt2001}.

Starting from this supersaturation mechanism proposed by \citet{StepienSchmitt2001}, we computed the expected variation in X-ray coronal emission with stellar rotation. Indicating with $\theta$ the stellar colatitude, then the local effective gravitational acceleration $g_{\rm eff}$ at the base of the convective envelope is
\[
g_{\rm eff}(\theta) = g\left(1-\frac{\omega^2R_{\rm core}\sin^2 \theta}{g}\right)
\]

\noindent
where $g = G M_{\rm core}/R_{\rm core}^2$ is the gravitational acceleration at the base of the convective envelope and $M_{\rm core}$ and $R_{\rm core}$ are the mass and radius of the inner radiative core. Considering that the energy flux at the base of the convective envelope $F(\theta)$ is proportional to $g_{\rm eff}(\theta)$, then the energy flux emerging through the convective envelope is expected to scale as $g_{\rm eff}^\alpha(\theta)$, with $\alpha$ being related to the dimension of the convective envelope \citep[with $\alpha\sim0.3$ for deep envelopes and $\alpha\sim1$ for shallow envelopes,][]{StepienSchmitt2001}. We assumed that the magnetic field emergence, and hence the local X-ray luminosity of coronal plasma, has the same dependence on stellar colatitude $\theta$ as $g_{\rm eff}^{\alpha}$. Therefore, from integrating over the whole stellar surface, the $L_{\rm X}$ should follow the relation

\[
L_{\rm X} = L_{\rm X0}\frac{1}{4{\rm \pi}}\int_{0}^{{\rm \pi}}\left(1-\frac{\omega^2R_{\rm core}\sin^2 \theta}{g}\right)^{\alpha}2{\rm \pi}\sin\theta\,{\rm d}\theta
\]

\noindent
where $L_{\rm X0}$ is again the saturated X-ray luminosity, which we assumed to be $10^{-3}$ with respect to $L_{\rm bol}$. We assumed for $M_{\rm core}$ and $R_{\rm core}$ the values predicted by \citet{SiessDufour2000} at an age of 13\,Myr. In the right panel of Fig.~\ref{figssmodel} we show the comparison between this predicted relations and the observed values, with $\alpha=1$ (a lower value for $\alpha$ would provide a shallower decrease of $L_{X}$ for increasing rotational velocity). From this plot we observe that in the predicted patterns $L_{\rm X}$ starts to reduce for $P_{\rm rot}\lesssim0.2$\,d, while the data clearly suggest that supersaturation starts to become effective for significantly higher $P_{\rm rot}$ (i.e. $P_{\rm rot}\sim1$\,d).

Evolution of stellar rotation predicts that the radiative core and the convective envelopes should have different rotation, with the inner radiative core rotating faster \citep[e.g.,][]{MacGregorBrenner1991,GalletBouvier2013}. We considered the hypothesis that the radiative core rotates
faster by $30\%$ than the convective envelope, as suggested by rotational evolutionary models of \citet{GalletBouvier2013}. The predicted patterns obtained (gray dashed lines in the right panel of Fig.~\ref{figssmodel}) indicate again that supersaturation caused by poleward migration occurs at $P_{\rm rot}$ shorter than that observed. To reconcile this discrepancy, a core rotational velocity higher by a factor $\sim 2-3$ than that of the convective envelope has to be assumed. The other possibility is that the polar updraft migration of active regions should act in a more efficient way at $P_{\rm rot}\sim0.5-1$\,d to be able to explain the observed supersaturation of the X-ray luminosity. There is another quite strong evidence against polar updraft migration as the mechanism responsible for supersaturation. The poleward shift of magnetic flux tubes, and hence of active regions, would reduce the surface filling factor of magnetically active regions,
but chromospheric activity indicators do not show any evidence of supersaturation \citep{MarsdenCarter2009,ChristianMathioudakis2011}, indicating that the surface of supersaturated stars, in comparison to that of saturated stars, is probably not depleted of active regions.

\subsection{Comparison with results obtained for younger or older stars}

With its evidence of non-saturation, saturation, and supersaturation, h~Per is the youngest cluster for which all the activity-rotation patterns have been observed. At an age of 13\,Myr, the stellar activity therefore starts to resemble that observed for MS stars and is usually interpreted as the manifestation of the $\alpha \Omega$ type dynamo. There are two main differences between h~Per members and younger stars: the internal structures, and the accretion status.

For the stellar internal structures, at an age of 13\,Myr all stars with $M_{\star}>0.5\,M_{\odot}$ have already developed
an inner radiative core, and the core contains more than half
of the stellar mass for stars with $M_{\star}>1.0\,M_{\odot}$ . At this age the internal structures are therefore analogous to those of MS stars, and in particular the $\alpha \Omega$ dynamo, which originates in the shell at the base of the convective envelope, can commence. Conversely, at an age of $1-3$\,Myr (the typical ages of very young PMS stars previously inspected),  stars with $1.0\,M_{\odot}<M_{\star}<1.4\,M_{\odot}$ are fully convective, or (depending on mass and age) have the convective envelope overwhelm the stellar structure in terms of
mass and volume, which means that the internal structure is different from that of MS stars. It is conceivable that in these very young stars the $\alpha \Omega$ dynamo efficiency is low because the base of the convective envelope
is buried very deeply in the stellar interior. Hence other dynamo mechanisms, like the turbulent dynamo \citep[e.g.,][]{DurneyDeYoung1993,ChabrierKuker2006} likely dominate. The coronal activity of very young stars is
therefore very likely regulated by a different dynamo mechanism.

Compared with very young stars, h Per members
also differ significantly because, at 13 Myr, they have already finished their accretion phase. Stars at a few Myr instead continue to accrete mass from their circumstellar disk. There are several
pieces of evidence that the X-ray emission of accreting stars is significantly different from that of non-accreting stars: accreting stars on average show hotter coronal plasma, lower X-ray luminosity, and a higher variability \citep[e.g.,][]{TsujimotoKoyama2002,StassunArdila2004,TelleschiGudel2007,FlaccomioMicela2012}. Several mechanisms have been proposed to explain the effect of accretion on coronal activity \citep[e.g.,][]{KastnerHuenemoerder2002,TelleschiGudel2007,BrickhouseCranmer2010}. However, this question is highly debated because the observed differences in the coronal emission of accreting and non-accreting stars could be explained in terms of different absorption suffered by the coronal emission \citep{FlaccomioMicela2010,FlaccomioMicela2012}. It is worth noting that even considering only very young stars with no accretion, the activity-rotation pattern in very young stars is unclear \citep[e.g.,][]{PreibischKim2005}. Therefore, even if accretion could affect magnetic activity of very young stars, it is probably not the cause of the observed difference of activity behavior between very young PMS and MS stars.

Assuming that centrifugal stripping is the mechanism causing supersaturation has important consequences when the results obtained from h~Per are extrapolated to younger or older stars. We expect that the $P_{\rm rot}$ for which centrifugal stripping occurs has to change with stellar age. PMS stars, evolving toward the MS phase, reduce their radii significantly. This does not change the co-rotation radius, that depends only on stellar mass and rotational period. However, radius contraction changes the location of the stellar surface, and hence of the corona, which on average approaches the stellar rotational axis. Assuming that coronal dimensions scale as the stellar radius and indicating with $\lambda$ the length of the largest coronal structures in units of stellar radii, centrifugal stripping becomes increasingly stronger for periods below the critical threshold of 

\[
P_{\rm rot} = 2{\rm \pi} \sqrt{ \frac{[(\lambda+1) R_{\star}]^3}{GM_{\star}} }
\]

\begin{figure}
\includegraphics[width=8.5cm]{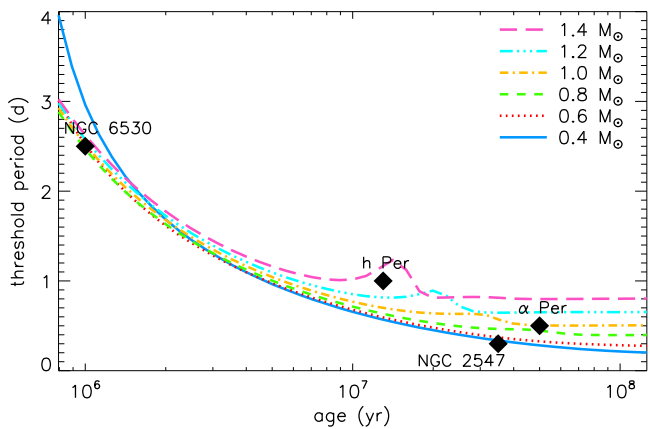}
\caption{Threshold value of $P_{\rm rot}$ for centrifugal stripping vs stellar age. These threshold values are computed assuming the \citet{SiessDufour2000} evolutionary models and that coronal structures extend up to 2 stellar radii above the stellar photosphere.}
\label{protvsage_ssmodel}
\end{figure}

\noindent with $M_{\star}$ and $R_{\star}$ being stellar mass and radius. Therefore, since $R_{\star}$ decreases during PMS evolution, shorter $P_{\rm rot}$ are needed for centrifugal stripping and hence supersaturation when stars contract. We show in Fig.~\ref{protvsage_ssmodel} the variation of this threshold period with stellar age for stars of different mass, assuming in all cases $\lambda=2$ during the entire PMS evolution. In addition to h~Per, hints of supersaturation were observed in a few clusters, and threshold periods were reported. In particular, \citet{HendersonStassun2012}, investigating a sample of $0.3-1.5\,M_{\odot}$ stars belonging to the \object{NGC~6530} cluster ($\sim1$\,Myr), found evidence of supersaturation for $P_{\rm rot}<2.5$\,d; \citet{JeffriesJackson2011} concluded that supersaturation occurs only for $P_{\rm rot}<0.3$\,d, considering a sample of $0.55-0.95\,M_{\odot}$ stars of \object{NGC~2547} ($\sim35$\,Myr); the oldest cluster (\object{$\alpha$~Per}, age $\sim50$\,Myr) was reported by \citet{RandichSchmitt1996}, who observed supersaturation only for G and K-type stars with $P_{\rm rot}<0.5$\,d. We report these observed thresholds in Fig.~\ref{protvsage_ssmodel}, concluding that they perfectly fit predictions based on centrifugal stripping. This agreement might also suggest that young stars are able to produce significant amounts of coronal structures with lengths of up to $\sim2\,R_{\star}$ during their entire PMS evolution, that is, from 1 to 50\,Myr.

\section{Conclusions}

We studied the activity-rotation relation in the young cluster \object{h~Per}, a $\sim13$\,Myr old cluster. This allowed us to investigate the processes in intermediate-age PMS stars, which, thanks to their age, show both fast and slow rotators, have completed the accretion process, and have developed an internal structure composed of an inner radiative core and an outer convective envelope. 

We found that solar-like ($1.0\,M_{\odot}<M_{\star}<1.4\,M_{\odot}$) h~Per members show different activity regimes, analogous to that observed in MS stars. This result makes h~Per the youngest cluster showing activity-rotation regimes. In particular, we clearly detected the supersaturation phenomenon for fast rotators, while slower rotators appear to be in the saturated or non-saturated regimes. Therefore when PMS stars develop a significant radiative core, their magnetic field production is most likely regulated by the $\alpha\Omega$ type dynamo, as occurs for MS stars.

The large numbers of fast rotating h~Per members allowed us to investigate the supersaturation phenomenon. We observed that supersaturation is better described by $P_{\rm rot}$ than $Ro$. Therefore the supersaturation phenomenon, at odds with other activity regimes that depend on the Rossby number, is not due to feedback effects that might inhibit the dynamo efficiency. Specifically, the observed patterns in the supersaturation regime strongly suggest that in fast rotators coronal emission is reduced because of centrifugal stripping, and that coronal structures have dimensions as large as $\sim2\,R_{\star}$ above the stellar surface. Moreover, the centrifugal stripping mechanism perfectly reproduces the observed evolution of the $P_{\rm rot}$ threshold for supersaturation with stellar age.

\begin{acknowledgements}
The scientific results reported in this paper are based on observations made by the {\it Chandra X-ray Observatory}.
\end{acknowledgements}

\bibliographystyle{aa} 
\bibliography{hper}

\Online

\scriptsize
\begin{longtable}{ccccr@{ $\pm$ }lr@{ $\pm$ }l}
\caption{X-ray source detected in the CHANDRA/ACIS-I observations of h~Per.} \\
\hline\hline
\label{tab:xsrc}
 X source ID & \multicolumn{1}{c}{RA}      & \multicolumn{1}{c}{DEC}     & \multicolumn{1}{c}{Pos. err.}   & \multicolumn{2}{c}{Count rate}                     & \multicolumn{2}{c}{Flux} \\
             & \multicolumn{1}{c}{(J2000)} & \multicolumn{1}{c}{(J2000)} & \multicolumn{1}{c}{($\arcsec$)} & \multicolumn{2}{c}{${\rm (10^{-5}\,cts\,s^{-1})}$} & \multicolumn{2}{c}{${\rm (10^{-8}\,ph\,s^{-1}\,cm^{-2})}$} \\
\hline
\endfirsthead
\caption{continued.} \\
\hline\hline
 X source ID & \multicolumn{1}{c}{RA}      & \multicolumn{1}{c}{DEC}     & \multicolumn{1}{c}{Pos. err.}   & \multicolumn{2}{c}{Count rate}                     & \multicolumn{2}{c}{Flux} \\
             & \multicolumn{1}{c}{(J2000)} & \multicolumn{1}{c}{(J2000)} & \multicolumn{1}{c}{($\arcsec$)} & \multicolumn{2}{c}{${\rm (10^{-5}\,cts\,s^{-1})}$} & \multicolumn{2}{c}{${\rm (10^{-8}\,ph\,s^{-1}\,cm^{-2})}$} \\
\hline
\endhead
\hline
\endfoot
\hline
\endlastfoot
    1 &           2:17:51.46 &           57:10:01.9 & 2.9 &      35.6 &   11 &     107.0 &   33 \\
    2 &           2:17:51.48 &           57:11:50.2 & 5.1 &      76.5 &   20 &     229.6 &   60 \\
    3 &           2:17:52.84 &           57:09:40.5 & 3.9 &      50.4 &   14 &     151.3 &   42 \\
    4 &           2:17:55.81 &           57:13:07.0 & 4.5 &      34.9 &   12 &     104.8 &   37 \\
    5 &           2:17:59.17 &           57:04:42.3 & 3.1 &      24.9 &    8 &      74.8 &   24 \\
    6 &          2:18:0 0.52 &           57:04:16.1 & 1.8 &      49.4 &    8 &     148.3 &   25 \\
    7 &          2:18:0 1.97 &           57:11:42.7 & 6.0 &      46.0 &   15 &     138.1 &   44 \\
    8 &          2:18:0 2.42 &           57:08:45.3 & 2.3 &      52.0 &    9 &     156.3 &   26 \\
\end{longtable}
\normalsize

\scriptsize
\begin{longtable}{cccccccr@{$\pm$}lr@{$\pm$}lccr@{$\pm$}l}
\caption{X-ray source identified as h~Per members.} \\
\hline\hline
\label{tab:xmemb}
 X memb & X src & Moraux2013$^a$ & Currie2010$^b$ & $\log L_{\rm X}^c$      & $m_{\rm V}^d$ & $m_{\rm I}^d$ & \multicolumn{2}{c}{$\log T_{\rm eff}$} & \multicolumn{2}{c}{$\log L_{\rm bol}/L_{\sun}$} & Period$^e$ & Mass$^e$           & \multicolumn{2}{c}{$\log \tau$} \\
 ID     & ID    & ID             & ID             &                         &               &               & \multicolumn{2}{c}{ }                  & \multicolumn{2}{c}{ }                           & (d)        & ${\rm (M_{\sun})}$ & \multicolumn{2}{c}{(d)}           \\
\hline
\endfirsthead
\caption{continued.} \\
\hline\hline
 X memb & X src & Moraux2013$^a$ & Currie2010$^b$ & $\log L_{\rm X}^c$      & $m_{\rm V}^d$ & $m_{\rm I}^d$ & \multicolumn{2}{c}{$\log T_{\rm eff}$} & \multicolumn{2}{c}{$\log L_{\rm bol}/L_{\sun}$} & Period$^e$ & Mass$^e$           & \multicolumn{2}{c}{$\log \tau$} \\
 ID     & ID    & ID             & ID             &                         &               &               & \multicolumn{2}{c}{ }                  & \multicolumn{2}{c}{ }                           & (d)        & ${\rm (M_{\sun})}$ & \multicolumn{2}{c}{d}           \\
\hline
\endhead
\hline
\endfoot
\hline
\multicolumn{15}{p{0.82\textwidth}}{
 
$^a$~Record ID of h~Per members listed in Table~2 of \citet{MorauxArtemenko2013}. 
$^b$~Record ID of h~Per members listed in Table~6 of \citet{CurrieHernandez2010}. 
$^c$~$L_{\rm X}$ are in ${\rm erg\,s^{-1}}$. 
$^d$~Apparent magnitudes in $V$ and $I_{\rm C}$ band are from \citet{CurrieHernandez2010}. 
$^e$~Period and mass values are form \citet{MorauxArtemenko2013}. 
\scriptsize
}
\endlastfoot
               1  &                 1 & $\cdot\cdot\cdot$ &              1056 & $ 30.64^{+0.12}_{-0.16}$ &        $         17.763$ &        $         16.274$ &                      $ 3.777$ & $ 0.010$ &          $\hspace{4.5pt} 0.46$ & $ 0.02$ & $\cdot\cdot\cdot$ & $\cdot\cdot\cdot$ &    \multicolumn{2}{c}{$\cdot\cdot\cdot$} \\
               2a &                 4 & $\cdot\cdot\cdot$ &             13614 & $ 30.51^{+0.13}_{-0.19}$ &        $         12.410$ &        $\cdot\cdot\cdot$ &                      $ 4.242$ & $ 0.106$ &          $\hspace{4.5pt} 3.01$ & $ 0.25$ & $\cdot\cdot\cdot$ & $\cdot\cdot\cdot$ &    \multicolumn{2}{c}{$\cdot\cdot\cdot$} \\
               2b &                 4 & $\cdot\cdot\cdot$ &                79 & $ 30.53^{+0.14}_{-0.21}$ &        $         12.444$ &        $         11.976$ &                      $ 4.404$ & $ 0.192$ &          $\hspace{4.5pt} 3.41$ & $ 0.53$ & $\cdot\cdot\cdot$ & $\cdot\cdot\cdot$ &    \multicolumn{2}{c}{$\cdot\cdot\cdot$} \\
               3  &                 7 & $\cdot\cdot\cdot$ &               772 & $ 30.59^{+0.12}_{-0.17}$ &        $         16.607$ &        $         15.414$ &                      $ 3.799$ & $ 0.014$ &          $\hspace{4.5pt} 0.63$ & $ 0.02$ & $\cdot\cdot\cdot$ & $\cdot\cdot\cdot$ &    \multicolumn{2}{c}{$\cdot\cdot\cdot$} \\
               4  &                 8 &               121 & $\cdot\cdot\cdot$ & $ 30.70^{+0.09}_{-0.12}$ &        $\cdot\cdot\cdot$ &        $\cdot\cdot\cdot$ &    \multicolumn{2}{c}{$\cdot\cdot\cdot$} &    \multicolumn{2}{c}{$\cdot\cdot\cdot$} & $           0.82$ & $           1.39$ &    \multicolumn{2}{c}{$\cdot\cdot\cdot$} \\
               5  &                10 &                45 &              1019 & $ 30.73^{+0.07}_{-0.09}$ &        $         17.739$ &        $         16.067$ &                      $ 3.701$ & $ 0.018$ &          $\hspace{4.5pt} 0.48$ & $ 0.03$ & $           2.55$ & $           1.33$ &                      $  2.33$ & $  0.25$ \\
\end{longtable}
\normalsize

\end{document}